\documentclass[aps,showpacs,preprintnumbers,amsmath,amssymb]{revtex4}
\usepackage{dcolumn}
\usepackage[dvips]{epsfig}

\begin{document}
\baselineskip=0.8 cm
\title{\bf Strong gravitational lensing by a rotating non-Kerr compact object}

\author{Songbai Chen\footnote{csb3752@hunnu.edu.cn}, Jiliang Jing
\footnote{jljing@hunnu.edu.cn}}

\affiliation{Institute of Physics and Department of Physics, Hunan
Normal University,  Changsha, Hunan 410081, People's Republic of
China \\ Key Laboratory of Low Dimensional Quantum Structures \\
and Quantum Control of Ministry of Education, Hunan Normal
University, Changsha, Hunan 410081, People's Republic of China}

\begin{abstract}
\baselineskip=0.6 cm
\begin{center}
{\bf Abstract}
\end{center}

We study the strong gravitational lensing in the background of a
rotating non-Kerr compact object with a deformed parameter
$\epsilon$ and an unbound rotation parameter $a$. We find that the
marginally circular stable orbit radius and the deflection angle
depend sharply on the parameters $\epsilon$ and $a$. For the case in
which the black hole is more prolate than a Kerr black hole, the
marginally circular photon orbit exists only in the regime
$\epsilon\leq\epsilon_{max}$ for a prograde photon. The upper limit
$\epsilon_{max}$ is a function of the rotation parameter $a$. As
$\epsilon>\epsilon_{max}$, the deflection angle of the light ray
very close to the naked singularity is a positive finite value,
which is different from that in the rotating naked singularity
described by Janis-Newman-Winicour metric. For the oblate black hole
and the retrograde photon, there does not exist such a threshold
value. Modeling the supermassive central object of the Galaxy as a
rotating non-Kerr compact object, we estimated the numerical values
of the coefficients and observables for gravitational lensing in the
strong field limit.

\end{abstract}

 \maketitle
\newpage
\section{Introduction}

The no-hair theorem \cite{noh} tells us that a neutral rotating
black hole in asymptotically flat and matter-free spacetime is
described completely by the Kerr metric with only two parameters,
the mass $M$ and the rotation parameter $a$,  which means that all
astrophysical black holes in our Universe should be Kerr black
holes. Since the current observational data show that there is still
lacking a definite proof for the existence of black holes in our
Universe, several potential methods have been proposed to test the
no-hair theorem by making use of observations including
gravitational waves from extreme mass-ratio inspirals
\cite{gwave,ga1,ga2,JGa} and the electromagnetic spectrum emitted by
the accreting disk around black holes \cite{ag1,CBa3}, and so on.
These techniques are based on spacetimes which deviate from the Kerr
metric by one or more parameters \cite{ga2,JGa,ag3,ag4}. Only if all
these deviations are measured to be zero, can we  ascertain that the
compact object is a Kerr black hole.

Motivated by examining the no-hair theorem, Johannsen and Psaltis
\cite{TJo} applied recently the Newman-Janis transformation
\cite{JNT} and constructed a Kerr-like black hole metric with a
deformed parameter $\epsilon$, which measures potential deviations
from the Kerr geometry. This rotating black hole possesses some
striking properties. For example, there are no restrictions on the
values of the rotation parameter $a$ and the deformation parameter
$\epsilon$, which means that it is possible that the rotation
parameter is larger than the mass of the black hole in this
spacetime. Moreover, the radius of horizon of the rotating non-Kerr
black hole depends on the Boyer-Lindquist polar angular coordinate
$\theta$, and the spacetime is free of closed timelike curves outside
of the outer horizon \cite{TJo}. Especially, as the deformation
parameter $\epsilon>0$, the black hole possesses two disconnected
spherical horizons for a high rotation parameter and has no horizon
for $a>M$. When $\epsilon<0$, the horizon always exists for the
arbitrary $a$, but the topology of the horizon becomes toroidal
\cite{CBa,CBa1}. The asymptotic behaviors of such a black hole in
the weak field approximation are the same as those of the usual Kerr
black hole in general relativity \cite{TJo}. These special
properties have attracted recently a great interest in the study of
the rotating non-Kerr black hole \cite{TJo, CBa,
CBa1,FCa,VCa1,TJo1,sc}.

Gravitational lensing is such a phenomenon resulting from the
deflection of light rays in a gravitational field. Like a natural
and large telescope, gravitational lensing can help us extract the
information about the distant stars which are too dim to be
observed. The strong gravitational lensing is caused by a compact
object with a photon sphere. When the photons pass close to the
photon sphere, the deflection angles become so large that an
observer would detect two infinite sets of faint relativistic images
on each side of the black hole, which are produced by photons that
make complete loops around the black hole before reaching the
observer. These relativistic images can provide us not only some
important signatures about black holes in the Universe, but also
profound verification of alternative theories of gravity in their
strong field regime \cite{Darwin,Vir,Vir1,Vir2,Vir3,Bozza2,Bozza3}.
Thus, the strong gravitational lensing is regarded as a powerful
indicator of the physical nature of the central celestial objects
and then has been studied extensively in various theories of gravity
\cite{Darwin,Gyulchev,Gyulchev1,Fritt,Bozza1,Eirc1,whisk,Bhad1,Song1,Song2,TSa1,AnAv,
Ls1,Kraniotis}.

The main purpose of this paper is to study the strong gravitational
lensing by a rotating non-Kerr compact object and to see whether it
can leave us the signature of the deformation parameter in the
deflection angle, the coefficients, and the observables for gravitational
lensing in the strong field limit. Moreover, we will explore how it
differs from the Kerr black hole lensing.

The paper is organized as follows: In the following section, we will
review briefly the rotating no-Kerr black hole metric proposed by
Johannsen and Psaltis \cite{TJo} to test the no-hair theorem in
the strong field regime, and then study the deflection angles for
light rays propagating in this background. In Sec.III, we study the
physical properties of the strong gravitational lensing by the
rotating non-Kerr compact object and probe the effects of the
deformation parameter on the marginally circular photon orbit
radius, the deflection angle, the coefficients, and the observables for
gravitational lensing in the strong field limit. We end the paper
with a summary.

\section{Rotating non-Kerr black hole spacetime and the deflection angles for light rays}

Let us now first review briefly the rotating no-Kerr black hole
metric, which was proposed by Johannsen and Psaltis \cite{TJo} to
test gravity in the strong field regime. Beginning with a deformed
Schwarzschild solution and applying the Newman-Janis transformation,
they constructed a deformed Kerr-like metric with three parameters:
the mass $M$, the rotation parameter $a$, and the deformation
parameter $\epsilon$. It is a stationary, axisymmetric, and
asymptotically flat spacetime. The line elements of the rotating
no-Kerr black hole in the standard Boyer-Lindquist coordinates can
be expressed as \cite{TJo}
\begin{eqnarray}
ds^2=g_{tt}dt^2+g_{rr}dr^2+g_{\theta\theta}d\theta^2+g_{\phi\phi}
d\phi^2+2g_{t\phi}dtd\phi, \label{metric0}
\end{eqnarray}
where
\begin{eqnarray}
g_{tt}&=&-\bigg(1-\frac{2Mr}{\rho^2}\bigg)(1+h),\;\;\;\;\;
g_{t\phi}=-\frac{2aMr\sin^2\theta}{\rho^2}(1+h),\nonumber\\
g_{rr}&=&\frac{\rho^2(1+h)}{\Delta+a^2h\sin^2\theta},\;\;\;\;\;\;\;\;\;\;\;\;\;\;\;
g_{\theta\theta}=\rho^2,\nonumber\\
g_{\phi\phi}&=&\sin^2\theta\bigg[r^2+a^2+\frac{2a^2Mr\sin^2\theta}{\rho^2}\bigg]
+\frac{a^2(\rho^2+2Mr)\sin^4\theta}{\rho^2}h,
\end{eqnarray}
with
\begin{eqnarray}
\rho^2=r^2+a^2\cos^2\theta,\;\;\;\;\;\;\;\;\;\;
\Delta=r^2-2Mr+a^2,\;\;\;\;\;\;\;\;\;\;h=\frac{\epsilon M^3
r}{\rho^4}.
\end{eqnarray}
The deformed parameter $\epsilon>0$ or $\epsilon<0$ corresponds to
the cases in which the compact object is more prolate or oblate than
a Kerr black hole, respectively. As $\epsilon=0$, the black hole is
reduced to the usual Kerr black hole in general relativity. The
position of the black hole horizon is defined by \cite{CBa,CBa1}
\begin{eqnarray}
\Delta+a^2h\sin^2\theta=0.
\end{eqnarray}
It is obvious that the radius of horizon is a function of the
Boyer-Lindquist polar angle $\theta$, which is quite a different
from that in the usual Kerr case. For the case $\epsilon>0$, one can
find that there exist two disconnected spherical horizons for high
spin parameters, but there is no horizon for $a>M$. However, for
$\epsilon<0$, it is shown  that the horizon never disappears for the
arbitrary $a$ and the shape of the horizon becomes toroidal
\cite{CBa,CBa1}. Moreover, one can find that it is free of closed
timelike curves and then causality is satisfied outside of the event
horizon \cite{TJo}.

Let us now study the strong gravitational lensing by a rotating
non-Kerr compact object. For simplicity, we here just consider that
both the observer and the source lie in the equatorial plane in the
rotating non-Kerr black hole spacetime (\ref{metric0}) and the whole
trajectory of the photon is limited on the same plane. With this
condition $\theta=\pi/2$, we obtain the reduced metric in the form
\begin{eqnarray}
ds^2&=&-A(x)dt^2+B(x)dx^2+C(x) d\phi^2-2D(x)dtd\phi, \label{metric1}
\end{eqnarray}
where we adopt to a new radial coordinate $x= r/2M$ and then the
metric coefficients become
\begin{eqnarray}
A(x)&=&\bigg(1-\frac{1}{x}\bigg)\bigg(1+\frac{\epsilon}{8x^3}\bigg),\\
B(x)&=&\frac{x^2(8x^3+\epsilon)}{8x^4(x-1)+a^2(8x^3+\epsilon)},\\
C(x)&=&x^2+a^2+\frac{a^2(8x^3+\epsilon+x\epsilon)}{8x^4},\\
D(x)&=&\frac{a(8x^3+\epsilon)}{8x^4}.
\end{eqnarray}
The null geodesics for the metric (\ref{metric1}) has the form
\begin{eqnarray}
\frac{dt}{d\lambda}&=&\frac{C(x)-JD(x)}{D(x)^2+A(x)C(x)},\label{u3}\\
\frac{d\phi}{d\lambda}&=&\frac{D(x)+JA(x)}{D(x)^2+A(x)C(x)},\label{u4}\\
\bigg(\frac{dx}{d\lambda}\bigg)^2&=&\frac{C(x)-2JD(x)-J^2A(x)}{B(x)C(x)[D(x)^2+A(x)C(x)]}.
\end{eqnarray}
where $\lambda$ is an affine parameter along the geodesics and $J$
is the angular momentum of the photon. In the background of a
rotating non-Kerr compact object (\ref{metric0}), one can obtain the
relation between the impact parameter $u(x_0)$ and the distance of
the closest approach of the light ray $x_0$ by the conservation of the
angular momentum of the scattering process
\begin{eqnarray}
u(x_0)=J(x_0)=\frac{8x_0^6+a^2(x_0+1)(8x^3_0+\epsilon)}{a(8x^3_0+\epsilon)
+x_0\sqrt{(8x^3_0+\epsilon)[8(x_0-1)x_0^4+a^2(8x^3_0+\epsilon)]}}.
\end{eqnarray}
Moreover, the deflection angle of the light becomes unboundedly large
as the closest distance of approach $x_0$ tends to the marginally
stable orbit radius $x_{ps}$ of the photon. In a stationary,
axially-symmetric metric, the equation of circular photon orbits
reads
\begin{eqnarray}
A(x)C'(x)-A'(x)C(x)+2J[A'(x)D(x)-A(x)D'(x)]=0.\label{root0}
\end{eqnarray}
The biggest real root external to the horizon of this equation
defines the marginally stable circular radius of photon
$x_{ps}=r_{ps}/2M$. For a rotating non-Kerr metric (\ref{metric0}),
the equation of circular photon orbits takes the form
\begin{eqnarray}
128x^9[x(2x-3)^2-8a^2]+32x^6(10x^3-27x^2+18x-18 a^2)\epsilon+2x^3
[x(5x-6)^2-48a^2]\epsilon^2-5 a^2\epsilon^3=0.\label{root}
\end{eqnarray}
Obviously, this equation depends on both the deformed parameter
$\epsilon$ and the rotation parameter $a$ of the compact object. The
presence of the deformed parameter $\epsilon$ makes the equation
more complex so that it is impossible to get an analytical form for
the marginally circular photon orbit radius in this case. In
Fig.(1), we present the variety of the marginally circular photon
orbit radius $x_{ps}$ with the deformed parameter $\epsilon$ and the
rotation parameter $a$ by solving Eq. (\ref{root}) numerically. It
is shown that the marginally circular photon orbit radius $x_{ps}$
decreases with the rotation parameter $a$ and the deformed parameter
$\epsilon$. This implies the variety of the marginally circular
photon orbit radius  $x_{ps}$ with the rotation parameter $a$ which
is similar to that in the Kerr black hole spacetime. Moreover, we also
find that the marginally circular photon orbit radius  $x_{ps}$
always exists for the case where the black hole is more oblate than a
Kerr black hole (i.e., $\epsilon<0$) or the case in which the black
hole rotates in the converse direction as the photon (i.e., $a<0$).
When the black hole is prolate ($\epsilon>0$) and it rotates in the
same direction as the photon ($a>0$), the marginally circular photon
orbit exists only in the regime $\epsilon<\epsilon_{max}$ for fixed
$a$. The value of $\epsilon_{max}$ is defined by the condition that
the marginally circular photon orbit is overlapped with the event
horizon (i.e., $x_{ps}=x_H$). In Table (I), we present the
largest value of the deformed parameter $\epsilon_{max}$ still
holding up the marginally circular photon orbit for fixed $a$, which
shows that it decreases with the parameter $a$.
\begin{figure}[ht]
\begin{center}
\includegraphics[width=6cm]{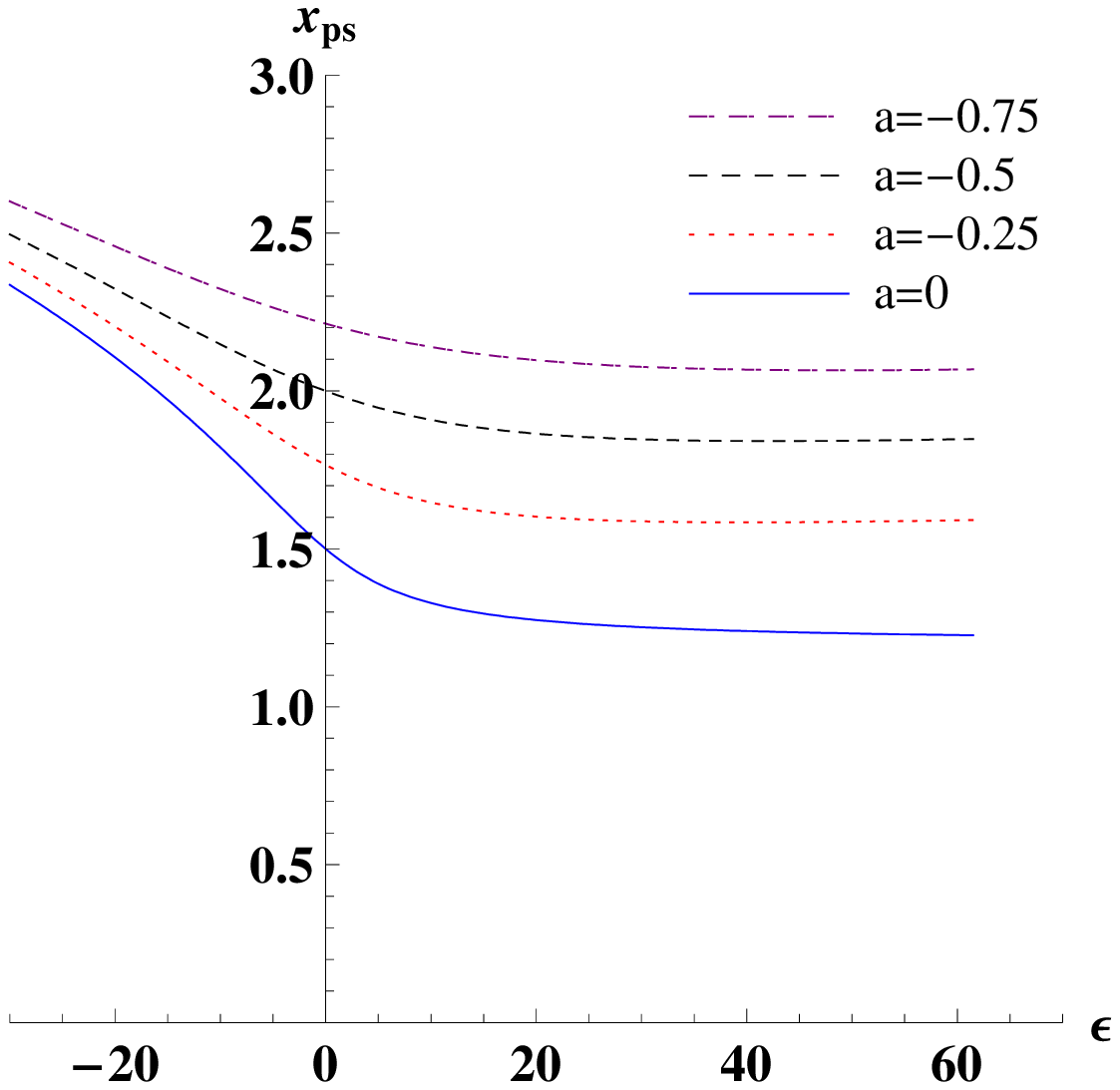}\;\;\;\;\includegraphics[width=6cm]{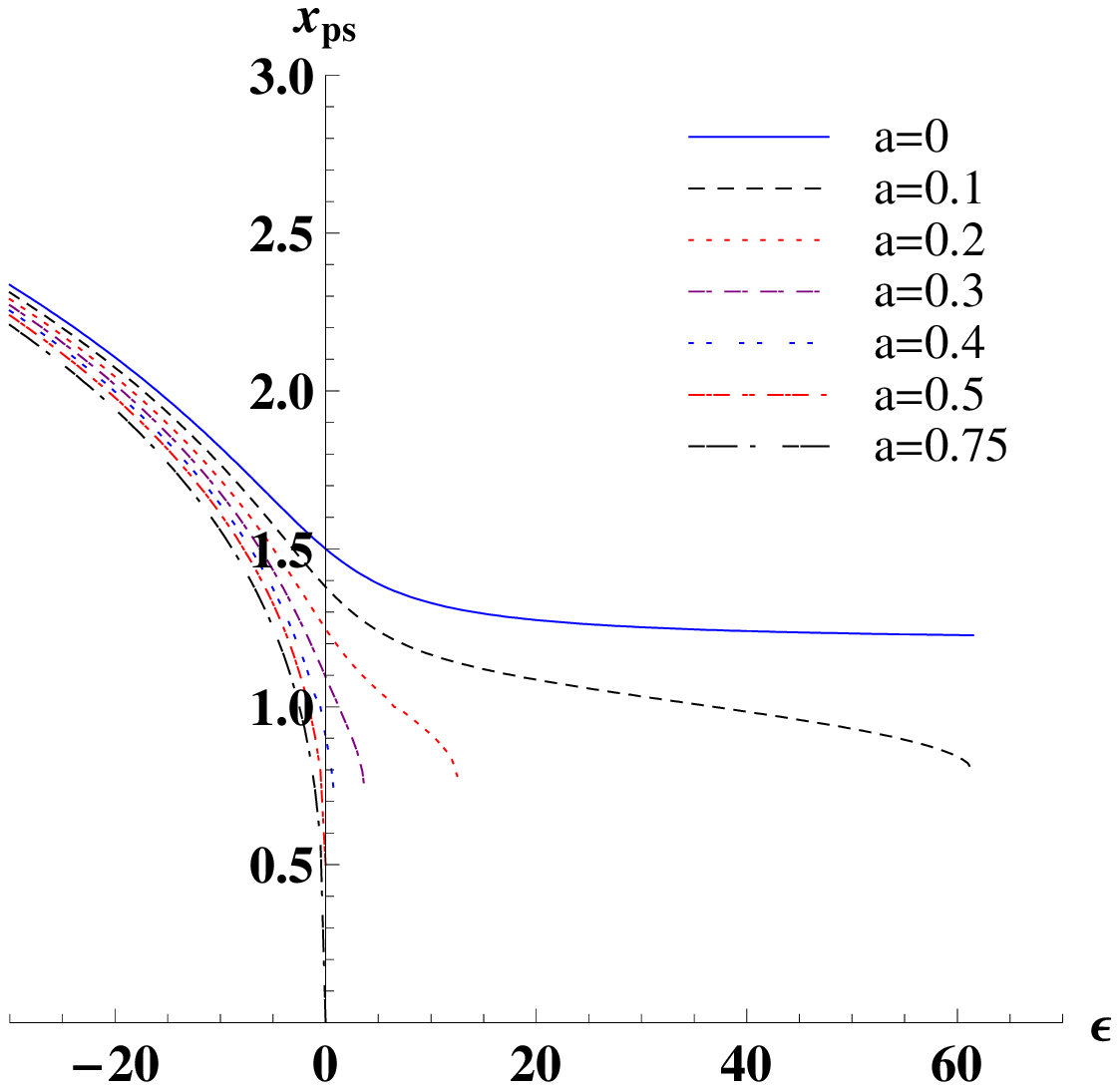}
\caption{Variety of the marginally circular orbit radius of the photon
with the deformed parameter $\epsilon$ for different $a$. Here, we
set $2M=1$.}
\end{center}
\end{figure}
\begin{table}[h]
\begin{center}
\begin{tabular}{c|cccccc}
\hline\hline  &&&&&&\\
$a$&0&0.1&0.2&0.3&0.4&0.5 \\
\hline
&&&&&&\\
$\epsilon_{max}$& $\infty$ &61.50&12.52&3.705&0.927 & 0
\\\hline
&&&&&&\\
$\text{lim}_{\epsilon\rightarrow\epsilon_{max}}x_{ps}$&
1.2&0.792&0.770&0.726&0.653&0.5 \\
\hline\hline
\end{tabular}
\end{center}
\label{tab1} \caption{The upper limit of the deformed parameter
$\epsilon$ still holding up the marginally circular photon orbit.}
\end{table}
When $\epsilon>\epsilon_{max}$, both the marginally circular
photon orbit and the event horizon vanish and then the singularity
is naked, which means that in this case the gravitational lensing
does not give relativistic images because of the nonexistence of
the marginally circular orbit of photon. This information implies
that the gravitational lensing by a rotating non-Kerr compact object
possesses some new special features, which could provide a
possibility to test the no-hair theorem in the strong field regime
in the near future.

The deflection angle for the photon coming from infinite in a
stationary, axially-symmetric spacetime, described by the metric
(\ref{metric1}) obeys \cite{Ein1}
\begin{eqnarray}
\alpha(x_{0})=I(x_{0})-\pi,
\end{eqnarray}
where $I(x_0)$ is given by
\begin{eqnarray}
I(x_0)=2\int^{\infty}_{x_0}\frac{\sqrt{B(x)|A(x_0)|}[D(x)+JA(x)]dx}{\sqrt{D^2(x)+A(x)C(x)}
\sqrt{A(x_0)C(x)-A(x)C(x_0)+2J[A(x)D(x_0)-A(x_0)D(x)]}}.\label{int1}
\end{eqnarray}
\begin{figure}[ht]
\begin{center}
\includegraphics[width=5cm]{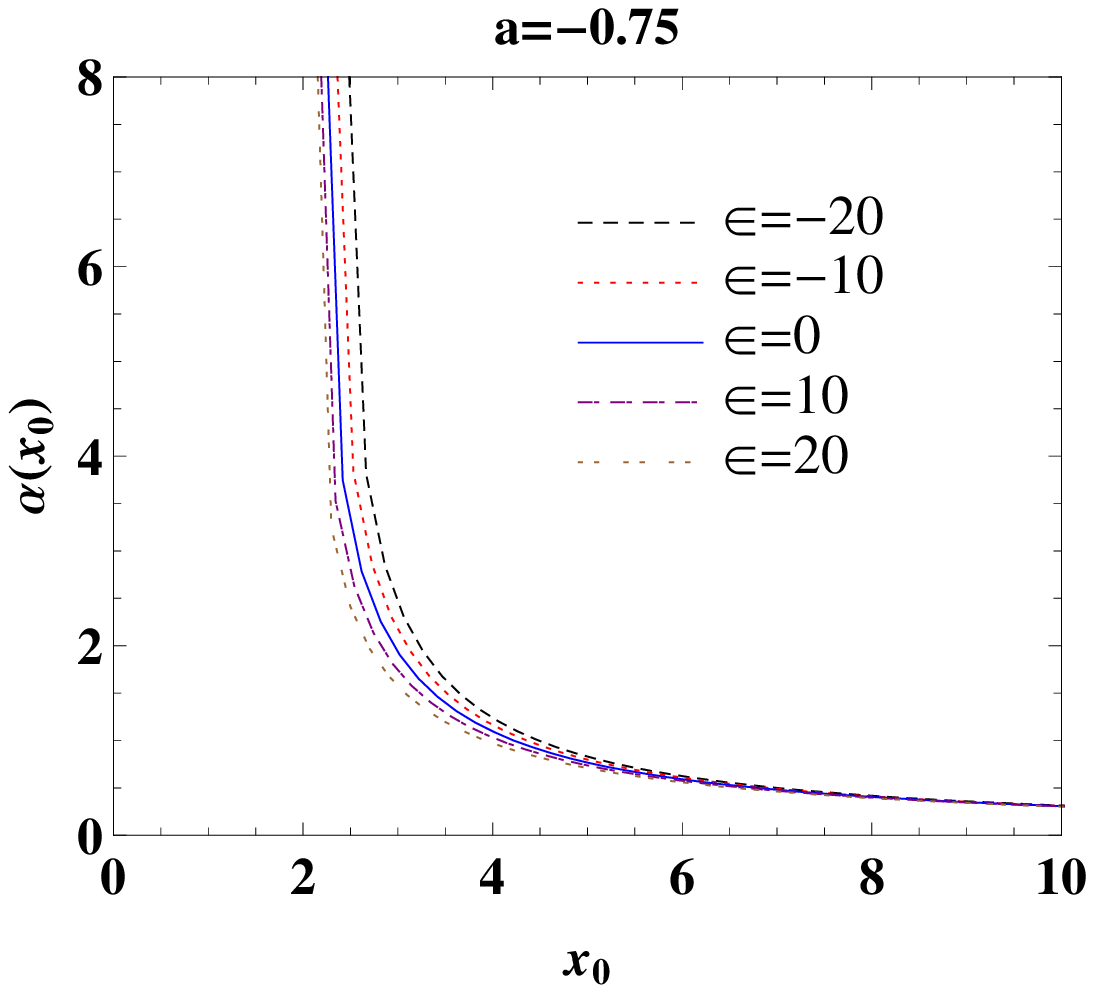}\;\includegraphics[width=5cm]{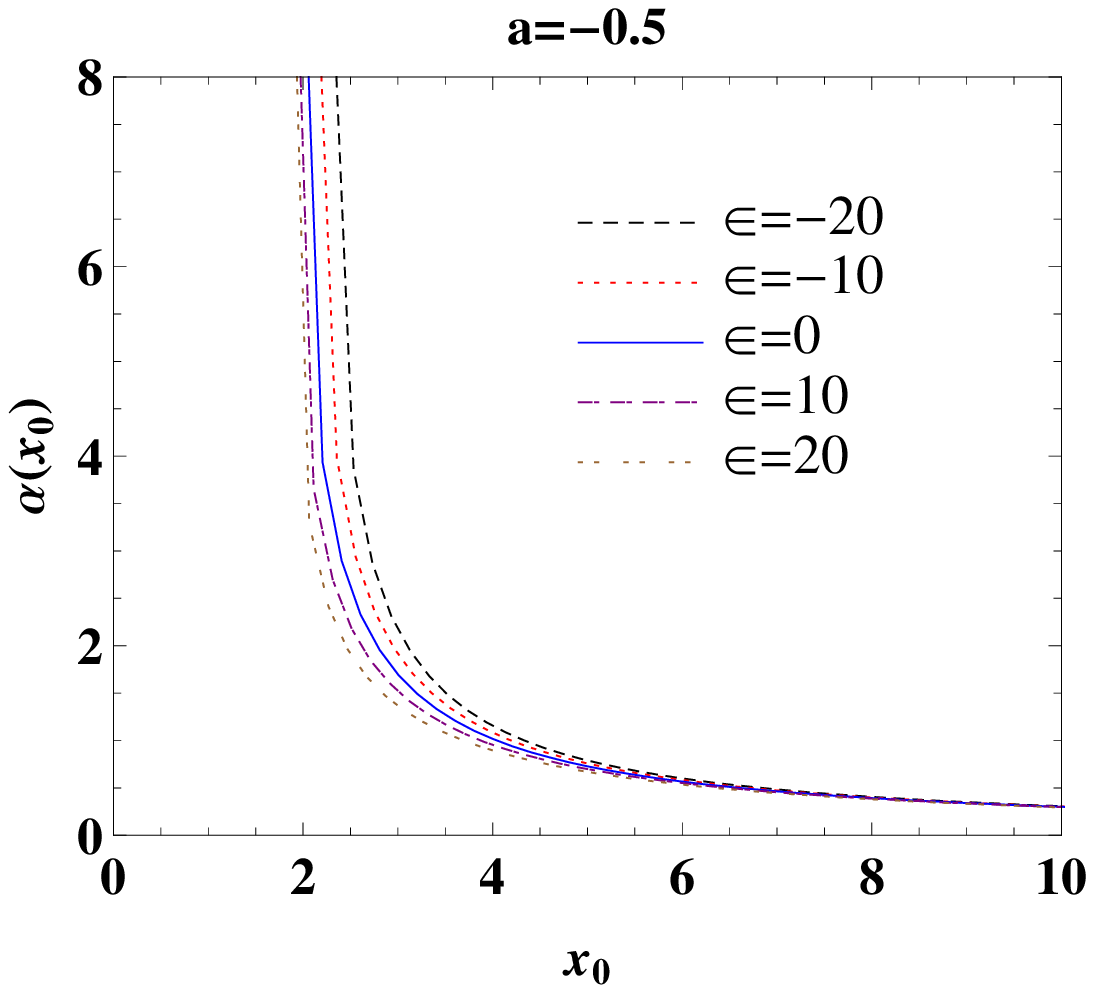}\;
\includegraphics[width=5cm]{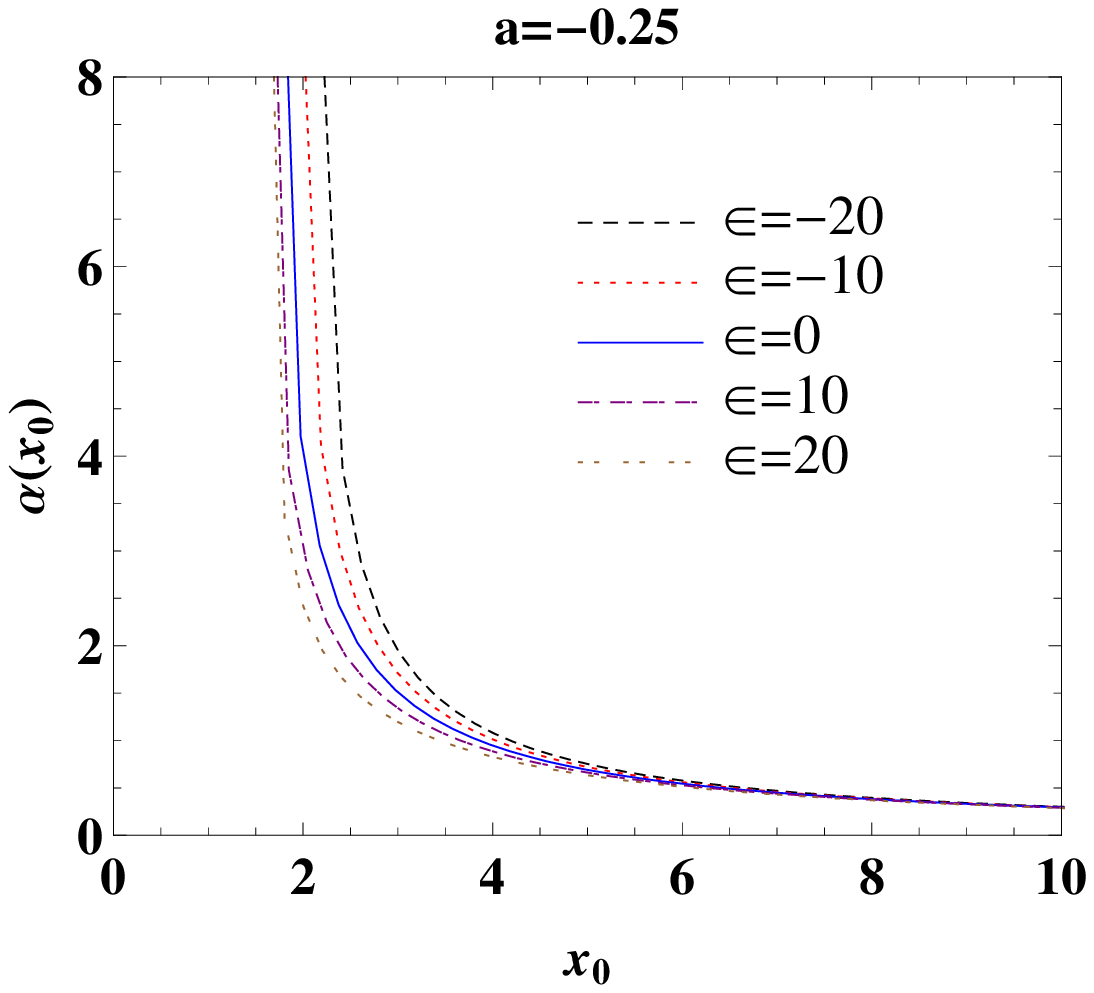}
\caption{Deflection angle $\alpha(x_0)$ as a function of the closest
distance of approach $x_0$ for angular momentum $a<0$. Here, we set
$2M=1$.}
\end{center}
\end{figure}
\begin{figure}[ht]
\begin{center}
\includegraphics[width=6cm]{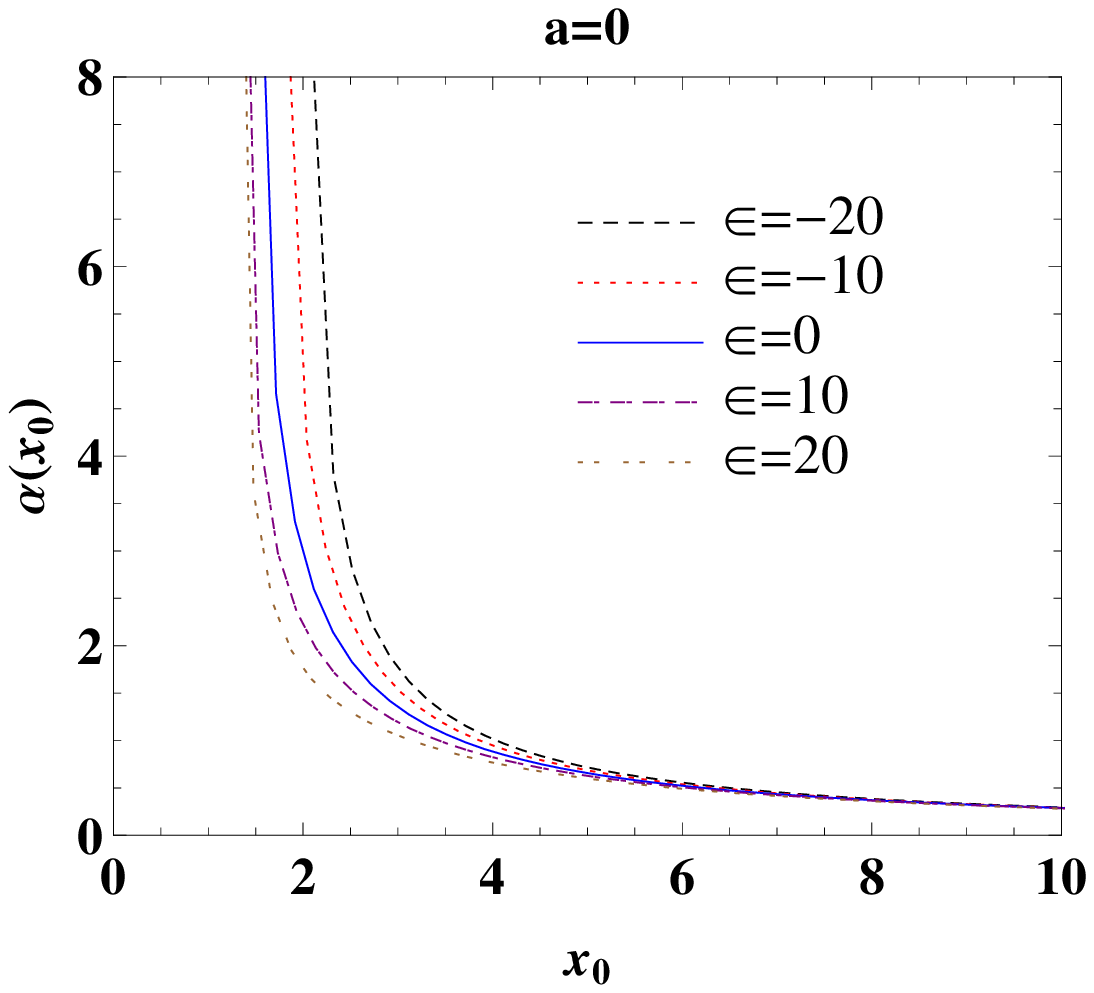}
\caption{Deflection angle $\alpha(x_0)$ as a function of the closest
distance of approach $x_0$ for angular momentum $a=0$. Here, we set
$2M=1$.}
\end{center}
\end{figure}
\begin{figure}[ht]
\begin{center}
\includegraphics[width=5cm]{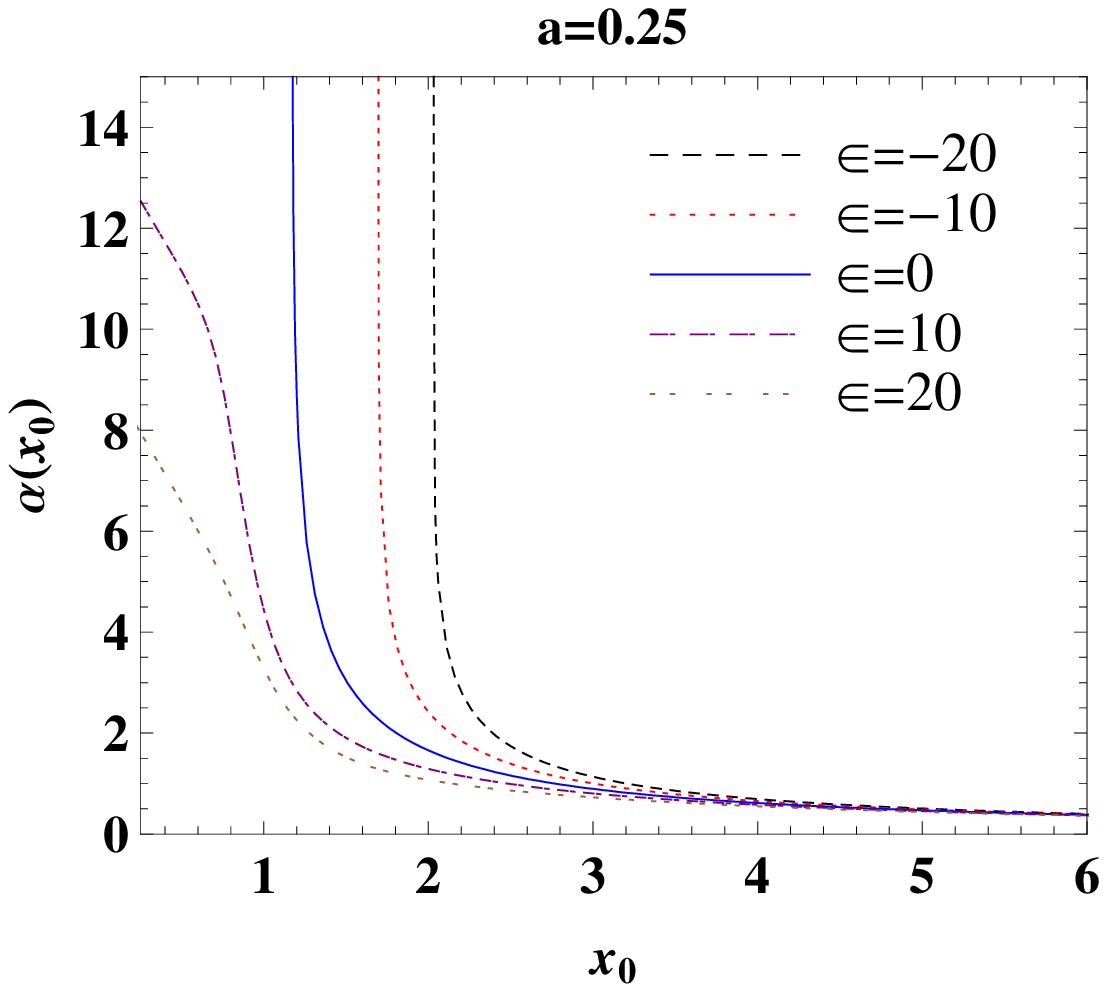}\;\includegraphics[width=5cm]{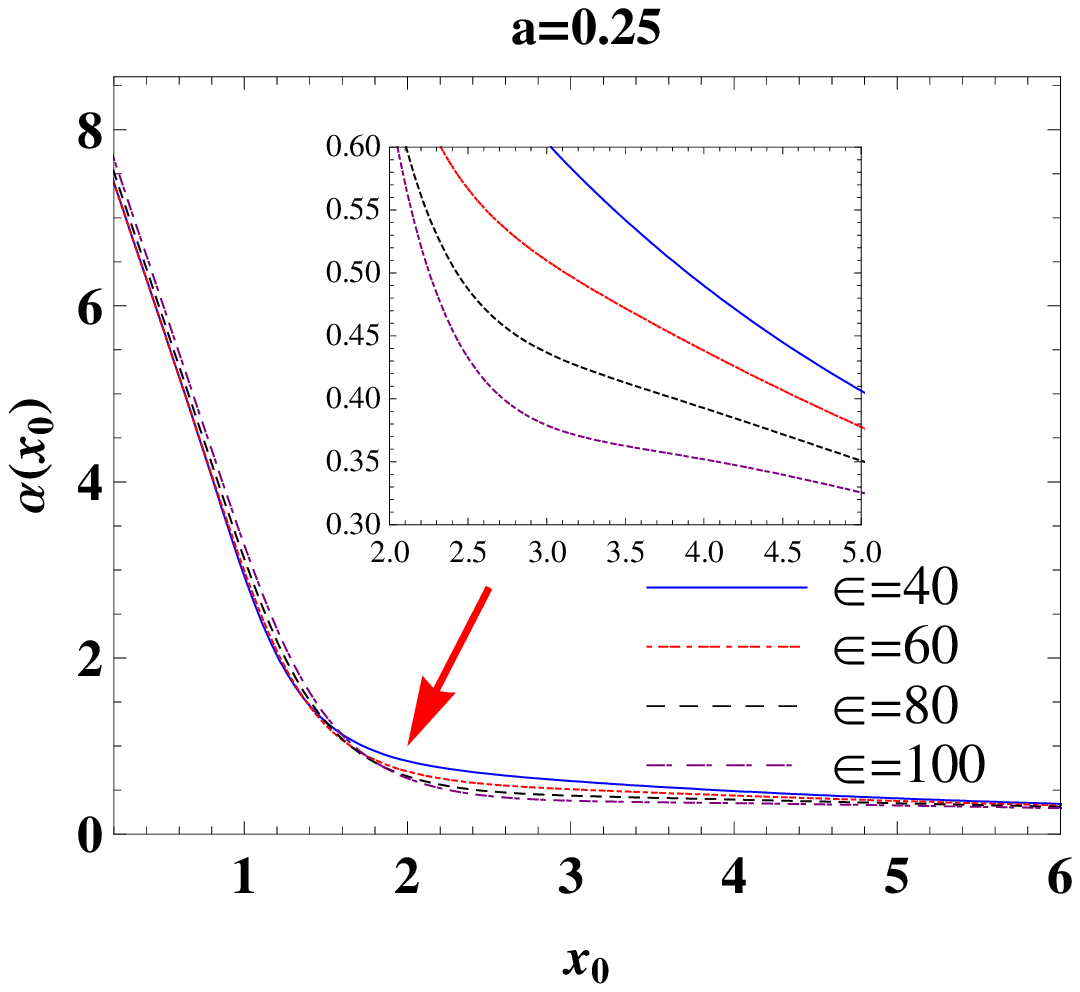}\;
\includegraphics[width=5cm]{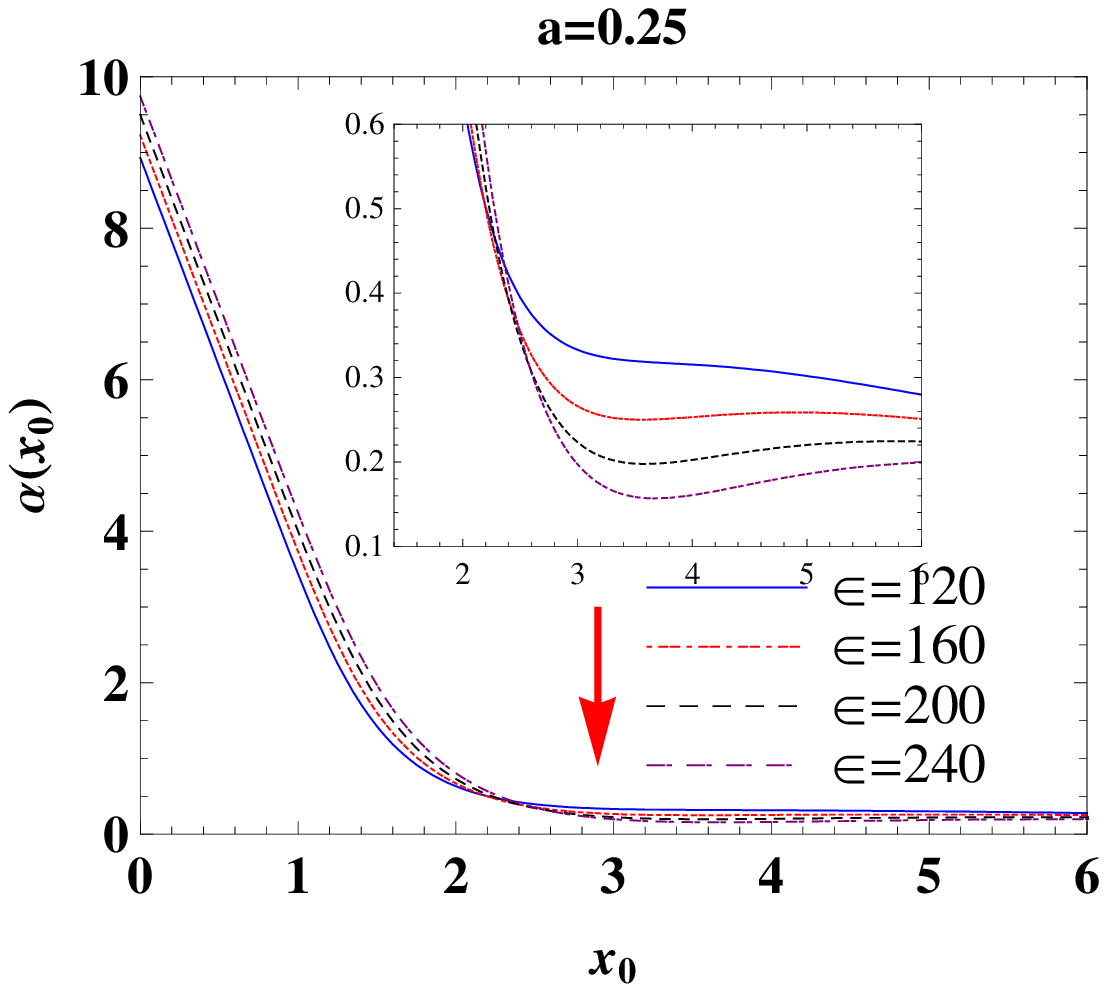}\\
\includegraphics[width=5cm]{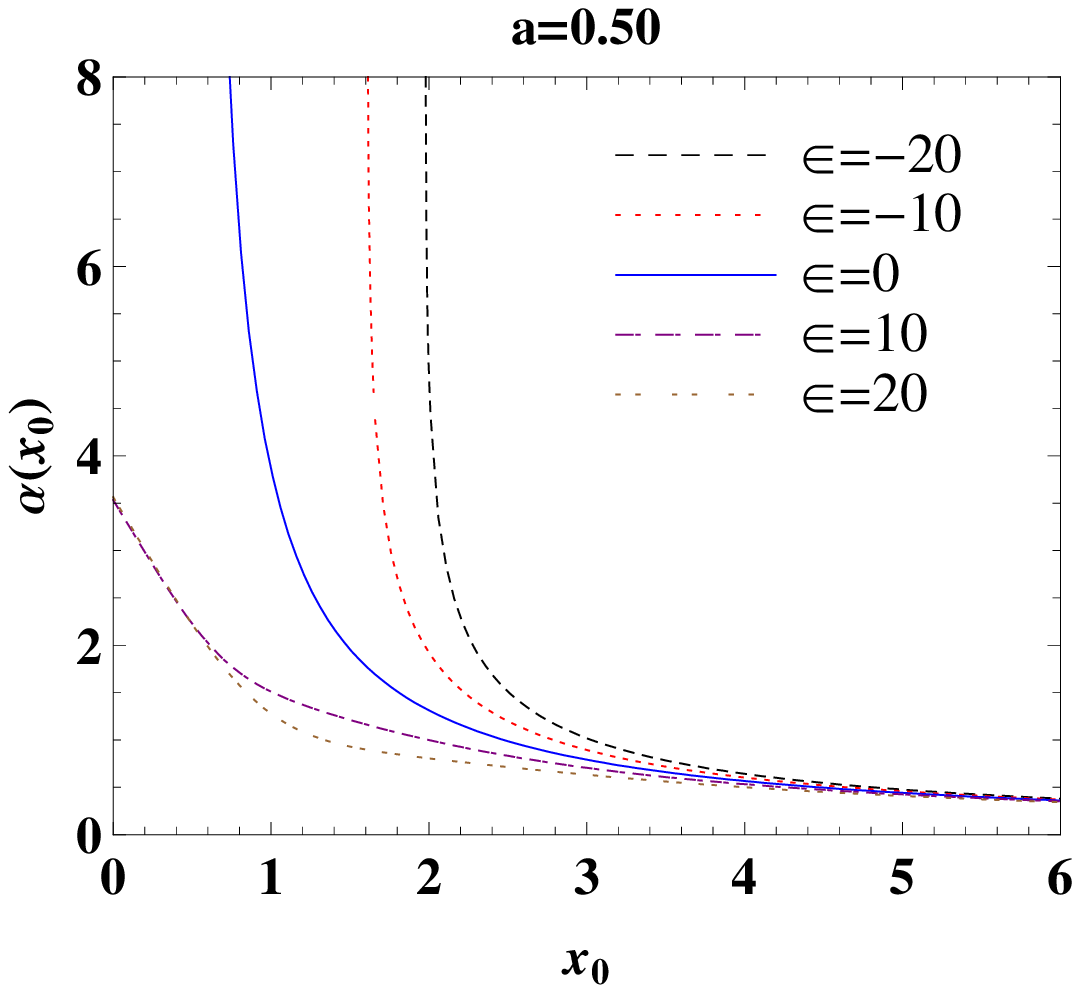}\;\includegraphics[width=5cm]{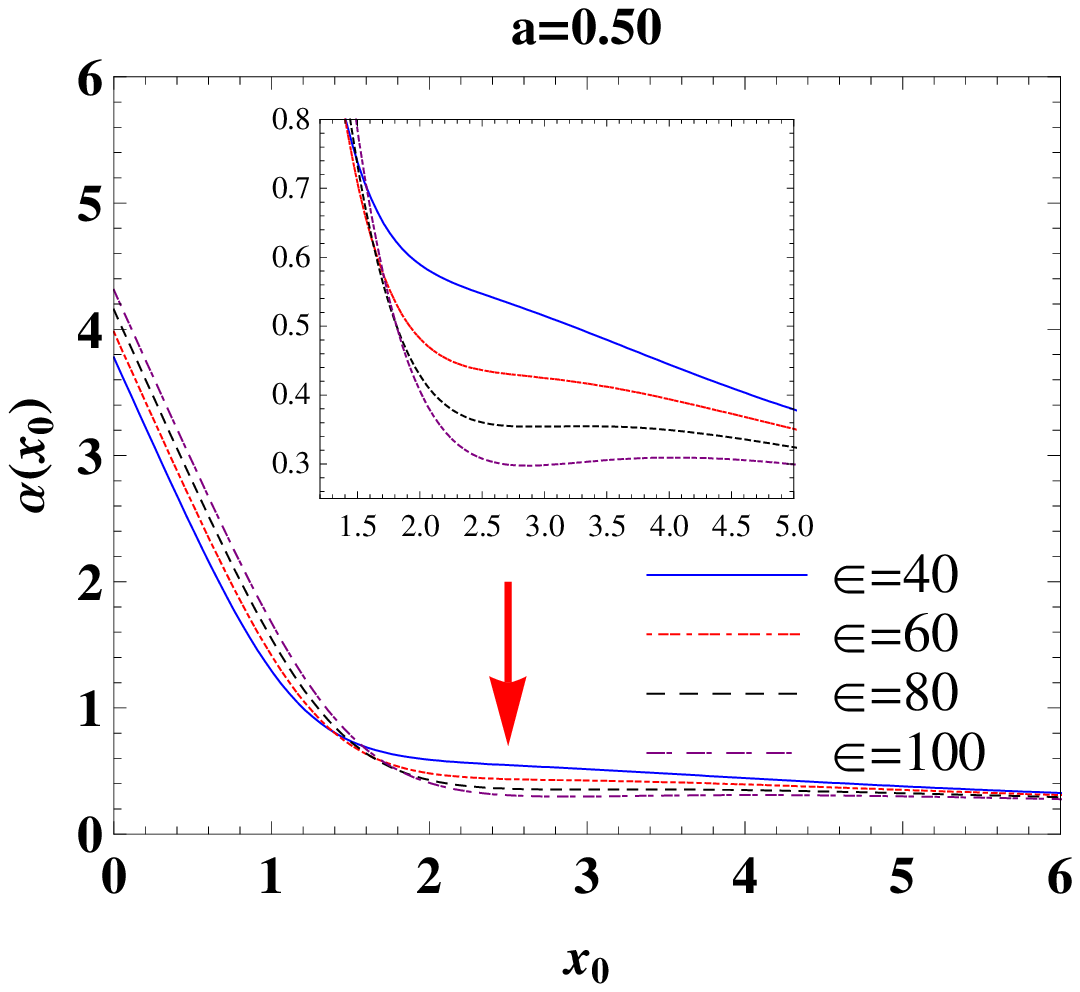}\;
\includegraphics[width=5cm]{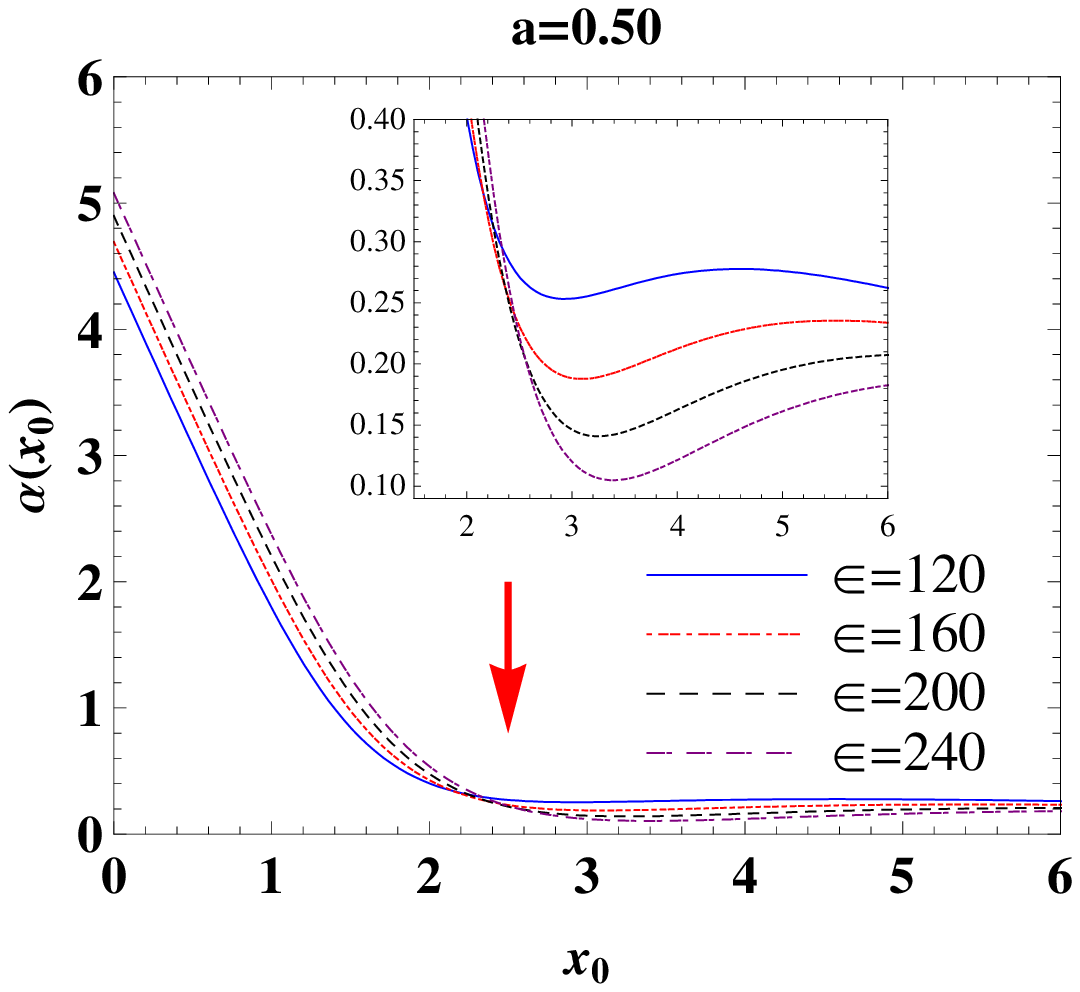}\\
\includegraphics[width=5cm]{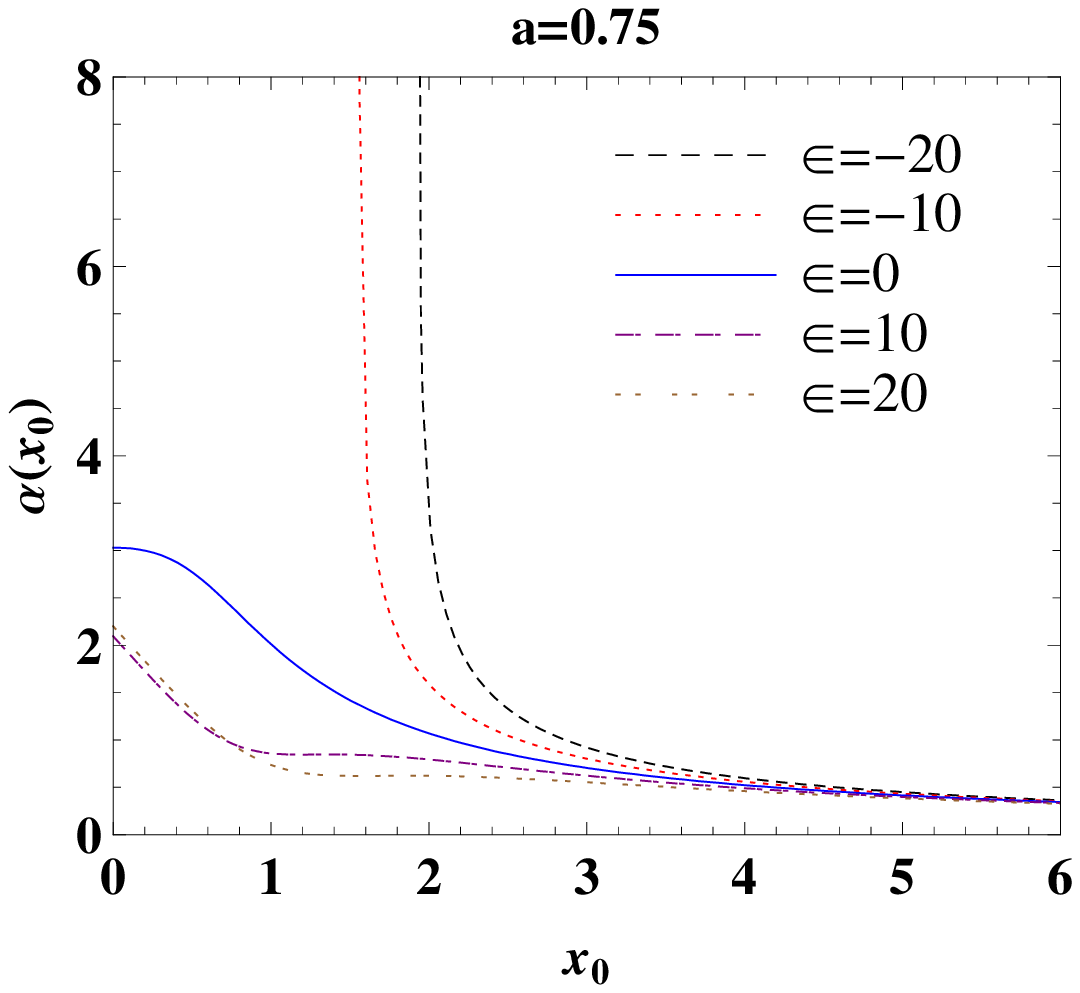}\;\includegraphics[width=5cm]{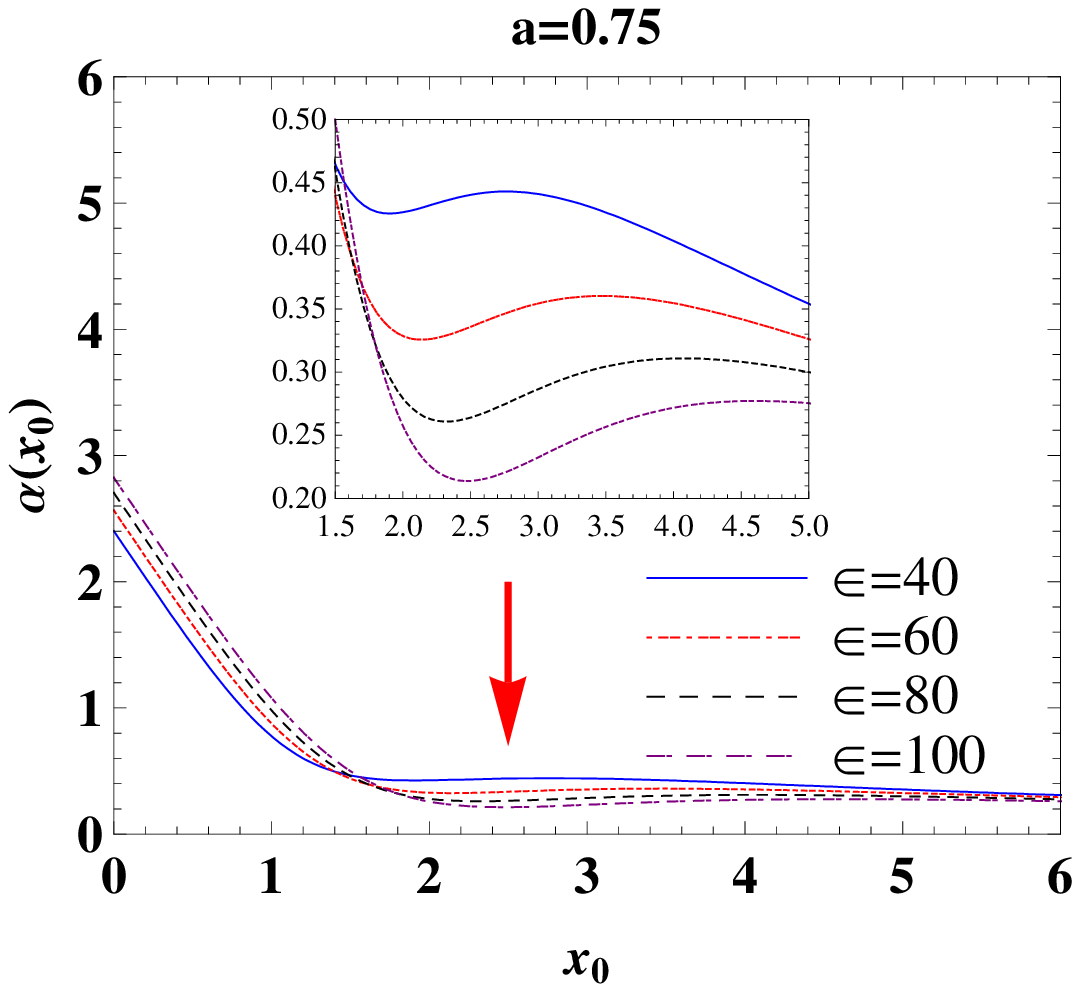}\;
\includegraphics[width=5cm]{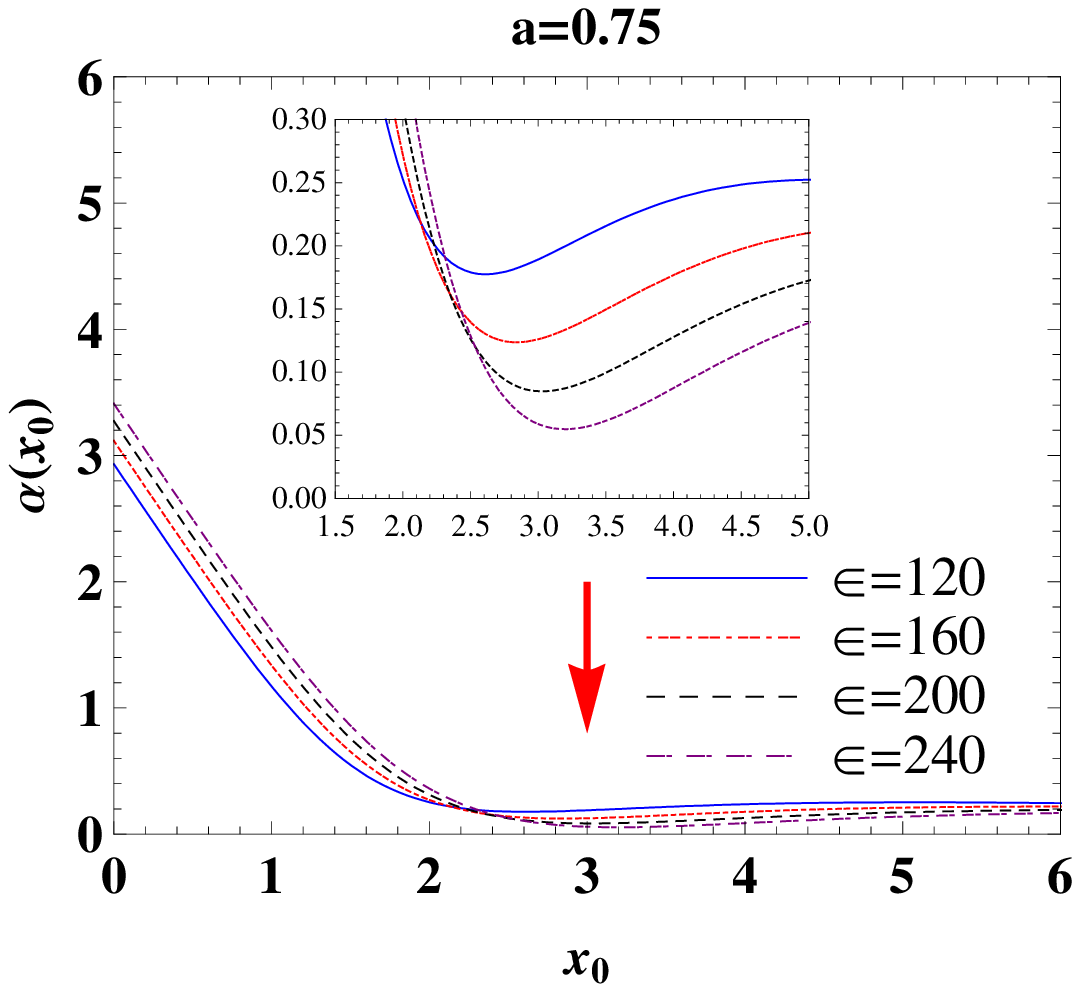}
\caption{Deflection angle $\alpha(x_0)$ as a function of the closest
distance of approach $x_0$ for angular momentum $a>0$. Here, we set
$2M=1$.}
\end{center}
\end{figure}
\begin{figure}[ht]
\begin{center}
\includegraphics[width=6cm]{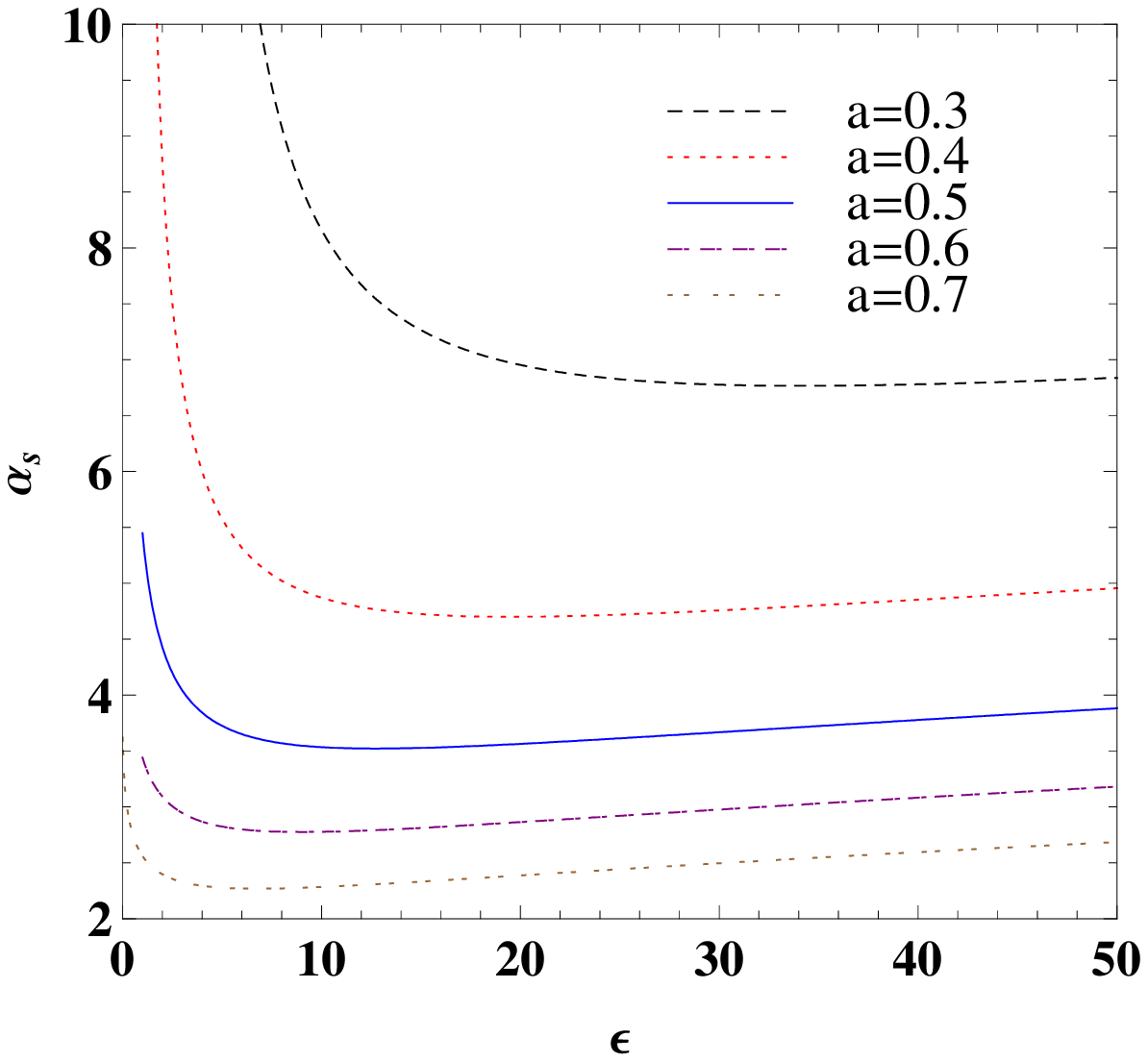}\;\;\includegraphics[width=6cm]{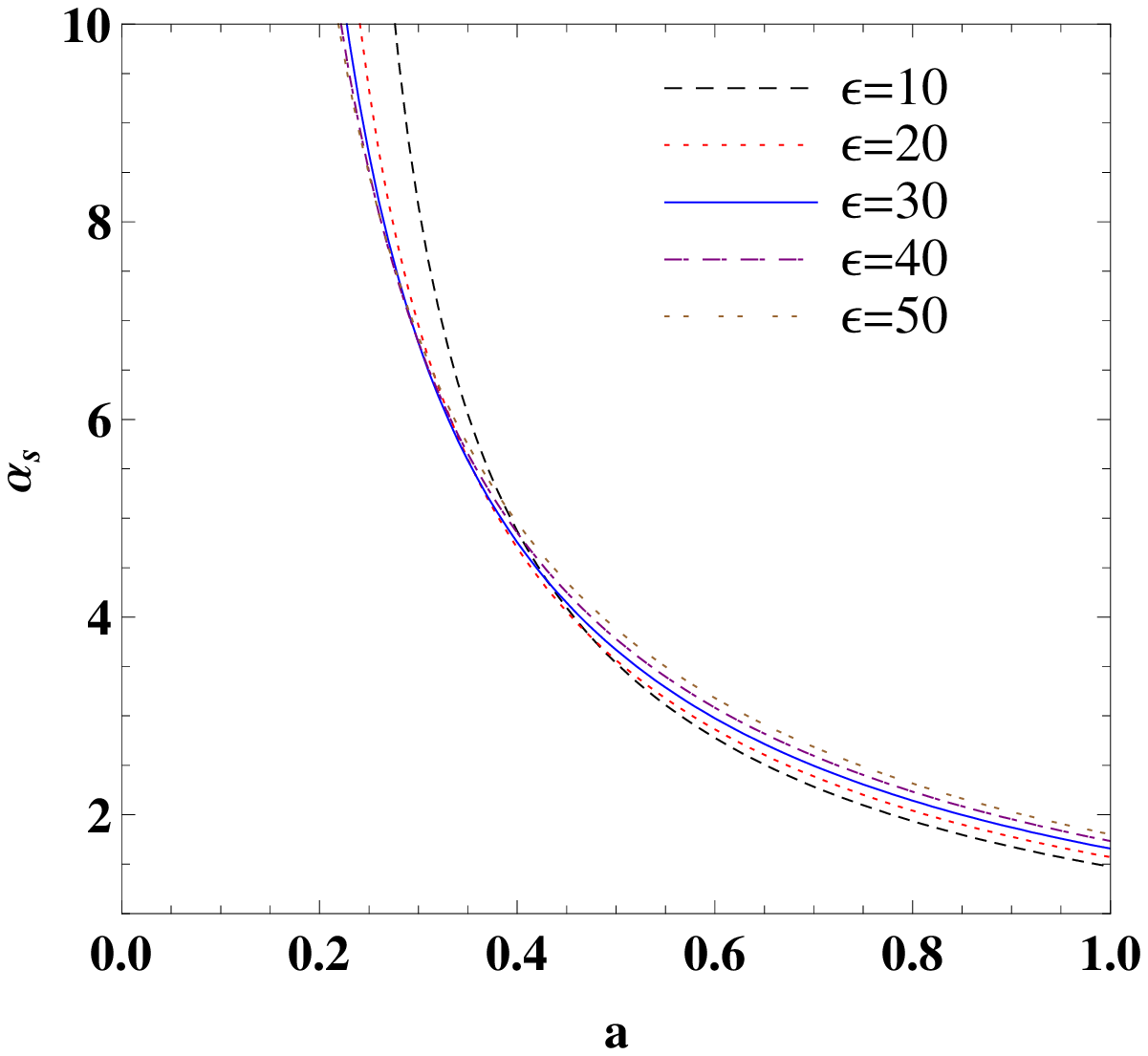}
\caption{Variety of the deflection angle $\alpha_s$ with the
deformed parameter $\epsilon$ and the angular momentum $a$ as the
light ray is very close to the singularity. Here, we set $2M=1$.}
\end{center}
\end{figure}
It is obvious that the deflection angle increases when parameter
$x_0$ decreases. For a certain value of $x_0$ the deflection angle
becomes $2\pi$, so that the light ray makes a complete loop around
the lens before reaching the observer. If $x_0$ is equal to the
radius of the marginally circular photon orbit $x_{ps}$, one can
find that the deflection angle becomes unboundedly large and the
photon is captured on a circular orbit. Let us now discuss the
behavior of the deflection angle for the lens described by a
rotating non-Kerr metric (\ref{metric0}).  From Figs. (2)-(4), one
can find that $\text{lim}_{x_0\rightarrow\infty}\alpha(x_{0})=0$ for
all values of $\epsilon$ and the angular momentum $a$, which is the
same as that in the Kerr case.  When there exists a marginally
circular photon orbit for the compact object, we find that the
deflection angle has similar qualitative behavior for the different
deformed parameter $\epsilon$ and it strictly increases with the
decreases of the impact parameter and becomes unboundedly large as
the closest distance of approach $x_0$ tends to the respective
marginally circular orbit radius $x_{ps}$ , i.e.,
$\text{lim}_{x_0\rightarrow x_{ps}}\alpha(x_{0})=\infty$ for $a<0$
and $a>0$, $\epsilon<\epsilon_{max}$. When there does not exist a
marginally circular photon orbit and the singularity is naked, we
find that the deflection angle of the light ray closing to the
singularity tends to a certain positive finite value $\alpha_s$,
i.e., $\text{lim}_{x_0\rightarrow 0}\alpha(x_{0})=\alpha_s$ for
$a>0$, $\epsilon>\epsilon_{max}$.  This is not surprising, because in
this case the event horizon has disappeared so that the photon could
not be captured by the rotating non-Kerr compact object. Moreover,
the similar behavior of the deflection angle was also obtained in
the Janis-Newman-Winicour spacetime with the naked singularity
\cite{Vir3,Gyulchev1}. The unique difference is that the finite
deflection angle is positive in the rotating non-Kerr spacetime
(\ref{metric0}) and is negative in the Janis-Newman-Winicour
spacetime \cite{Vir3,Gyulchev1}, which could be explained by the fact
that the difference in the spacetime structures between the rotating
non-Kerr spacetime and the Janis-Newman-Winicour spacetime leads to
the different properties of the naked singularities. This implies
that the gravitational lensing could provide a possible way to
distinguish the singularities in various spacetimes. The change of
$\alpha_s$ with the deformed parameter $\epsilon$ and the angular
momentum $a$ is shown in Fig. (5), which tells us that it decreases
monotonically with $a$ and first decreases and then increases with
the deformed parameter $\epsilon$. Moreover, we also note that for
the larger deformed parameter $\epsilon$ and angular momentum $a$
there exists a minimum for the deflection angle, which depends on
the parameters $\epsilon$ and $a$.

\section{Strong gravitational lensing in the rotating non-Kerr spacetime}

In this section we will study the gravitational lensing by the
rotating non-Kerr compact object with the marginally circular photon
orbit and then probe the effects of the deformed parameter
$\epsilon$ on the coefficients and the lensing observables in the
strong field limit.

In order to find the behavior of the deflection angle very close to
the marginally circular photon orbit, we adopt the evaluation
method for the integral (\ref{int1}) proposed by Bozza
\cite{Bozza2}. The divergent integral (\ref{int1}) is first split
into the divergent part $I_D(x_0)$ and the regular one $I_R(x_0)$,
and then both of them are expanded around $x_0=x_{ps}$ and with
sufficient accuracy are approximated with the leading terms. This
technique has been widely used in studying the strong
gravitational lensing of various black holes
\cite{Vir,Vir1,Vir2,Vir3,Bozza2,Bozza3,Gyulchev,Gyulchev1,Fritt,Bozza1,Eirc1,
whisk,Bhad1,Song1,Song2,TSa1,AnAv,Ls1,Darwin}. Let us now to define
a variable
\begin{figure}[ht]
\begin{center}
\includegraphics[width=5.2cm]{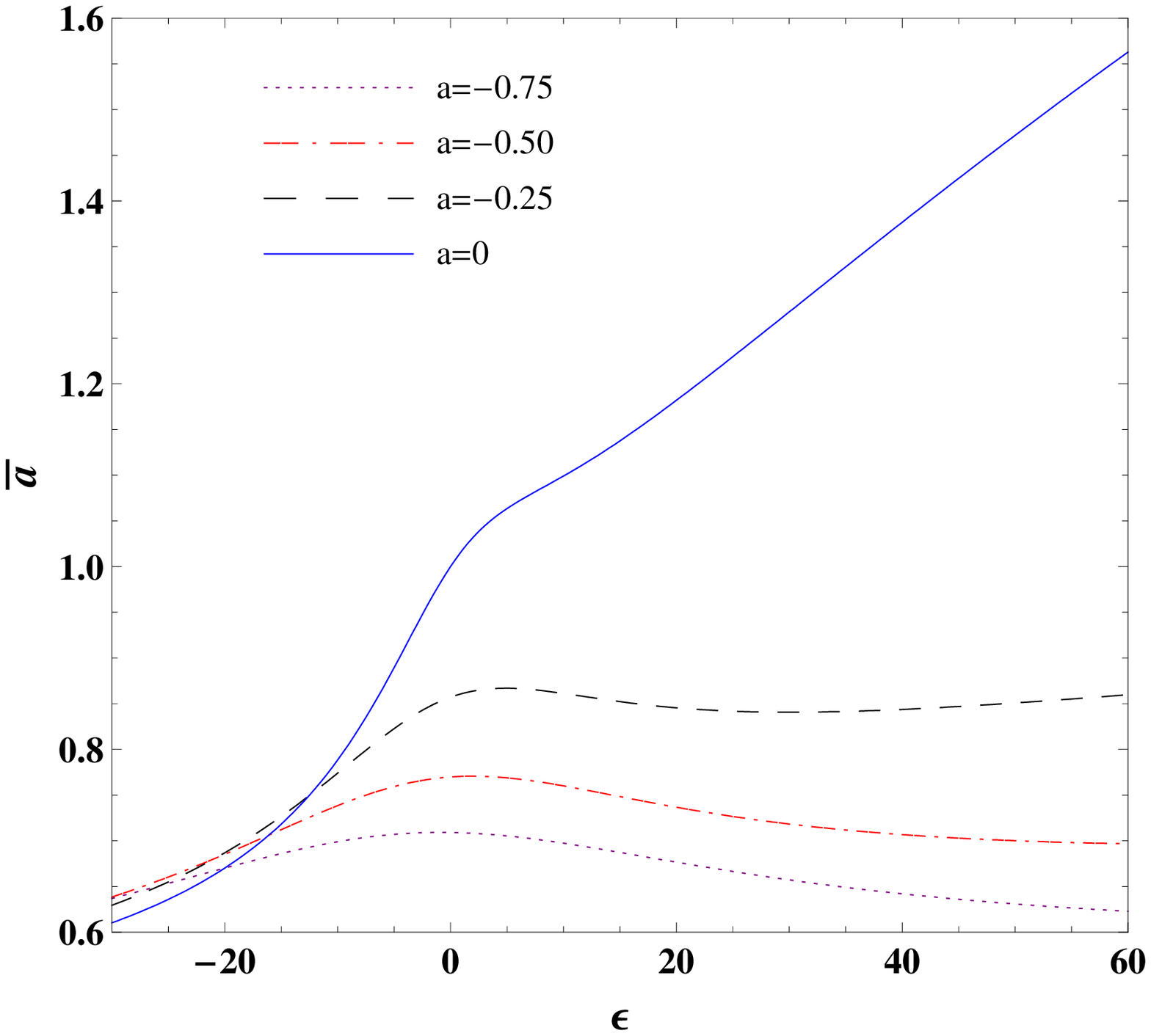}\;\;\includegraphics[width=5.2cm]{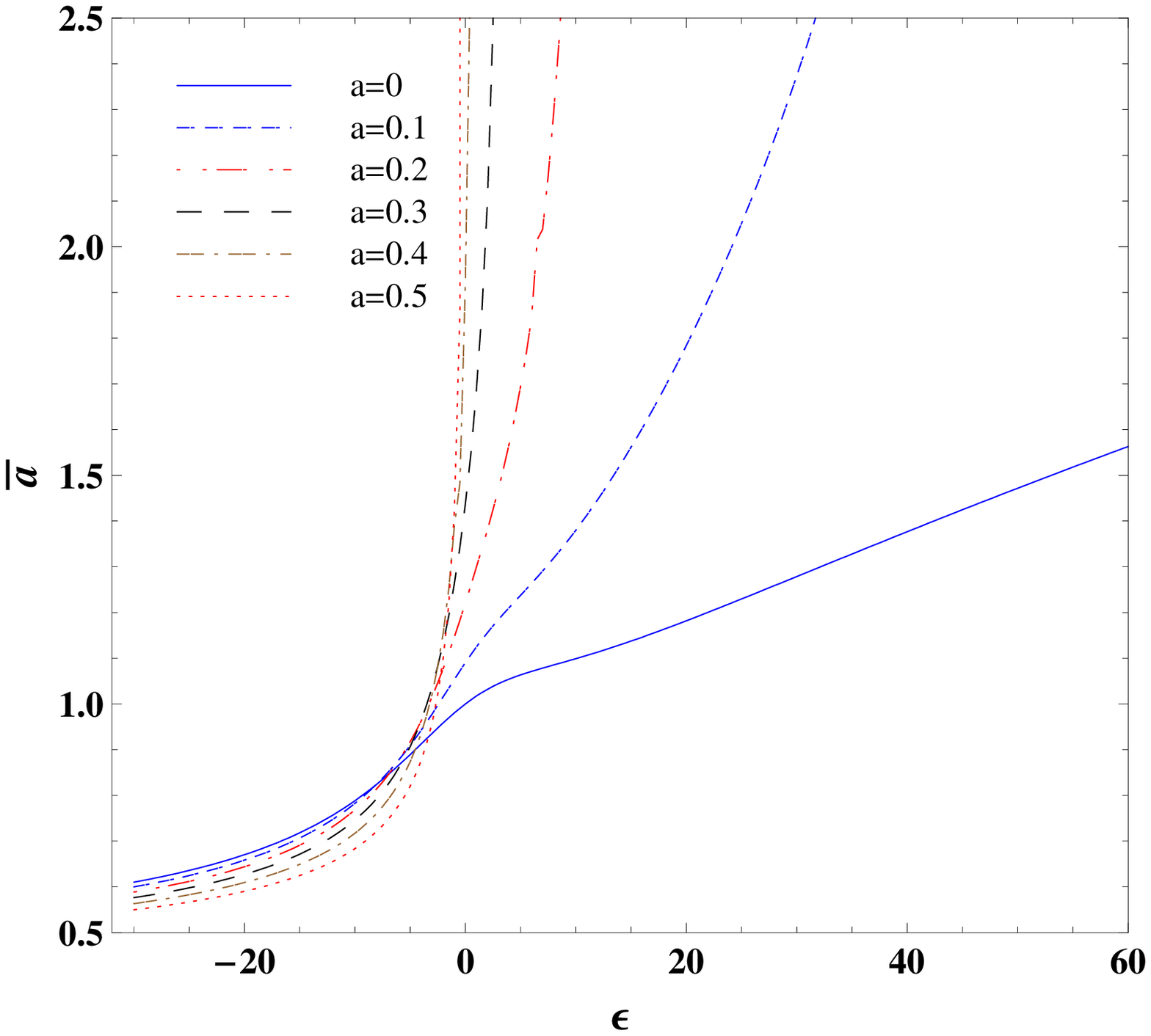}
\;\;\includegraphics[width=5.2cm]{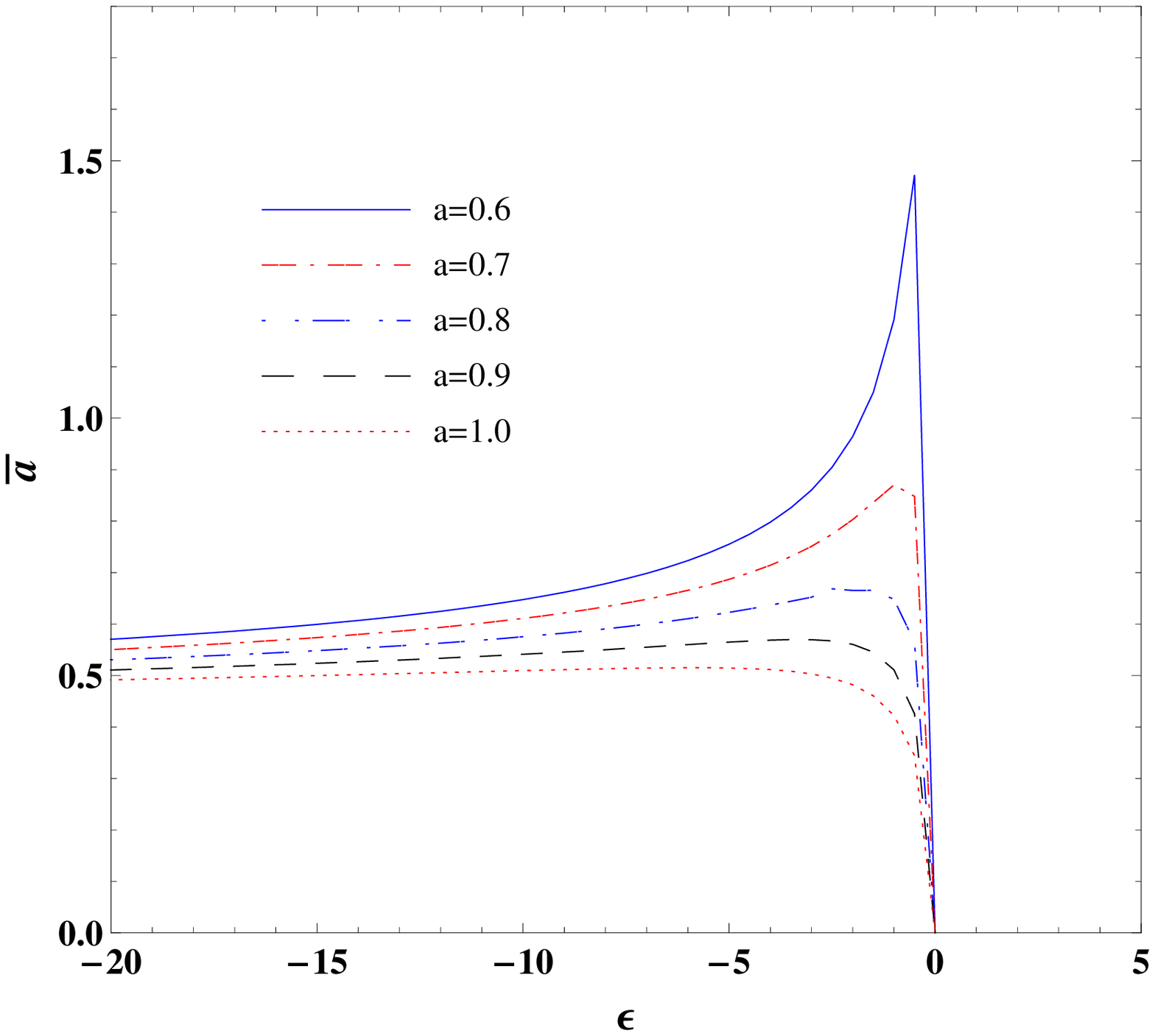}\\
\includegraphics[width=5.2cm]{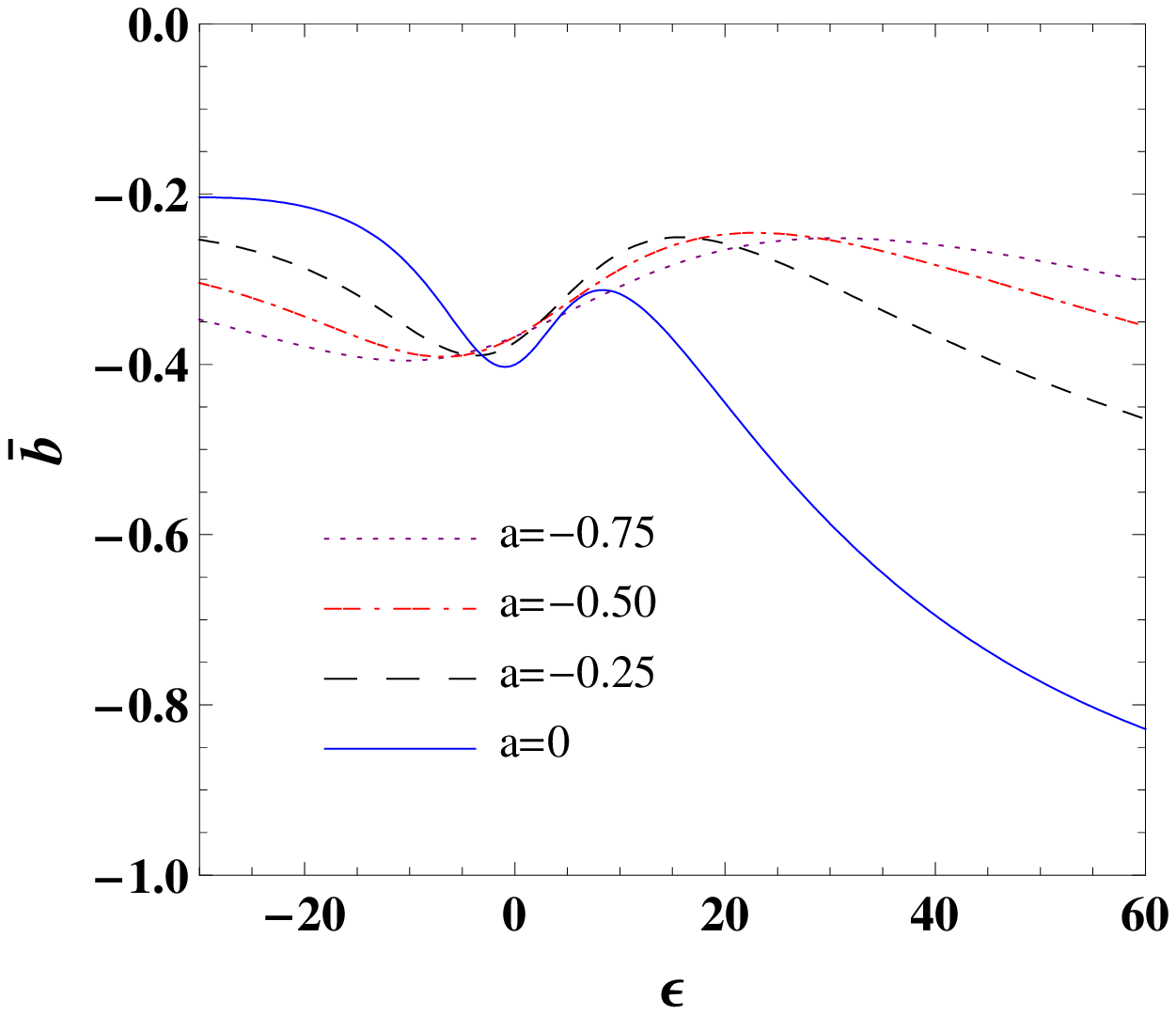}\;\;\includegraphics[width=5.8cm]{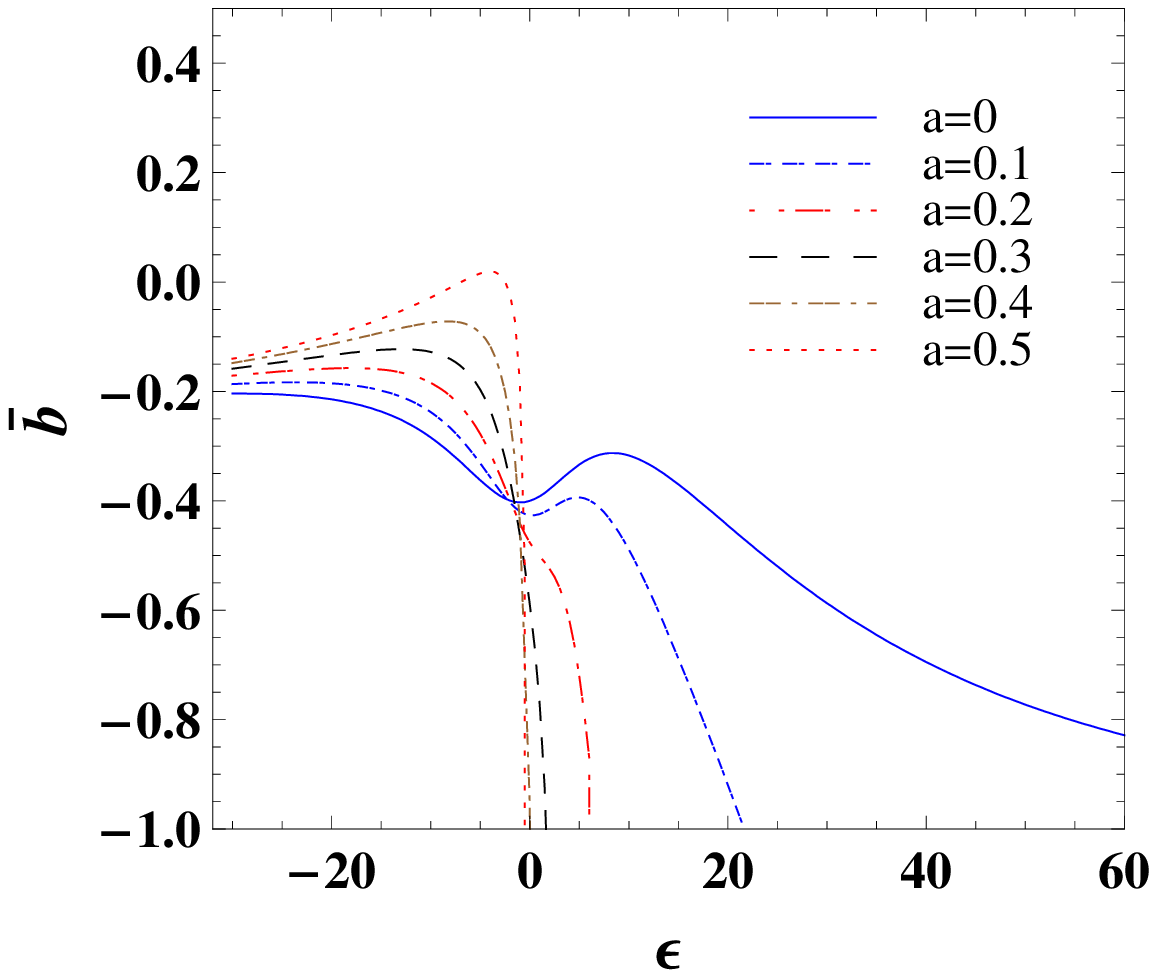}
\;\;\includegraphics[width=5.0cm]{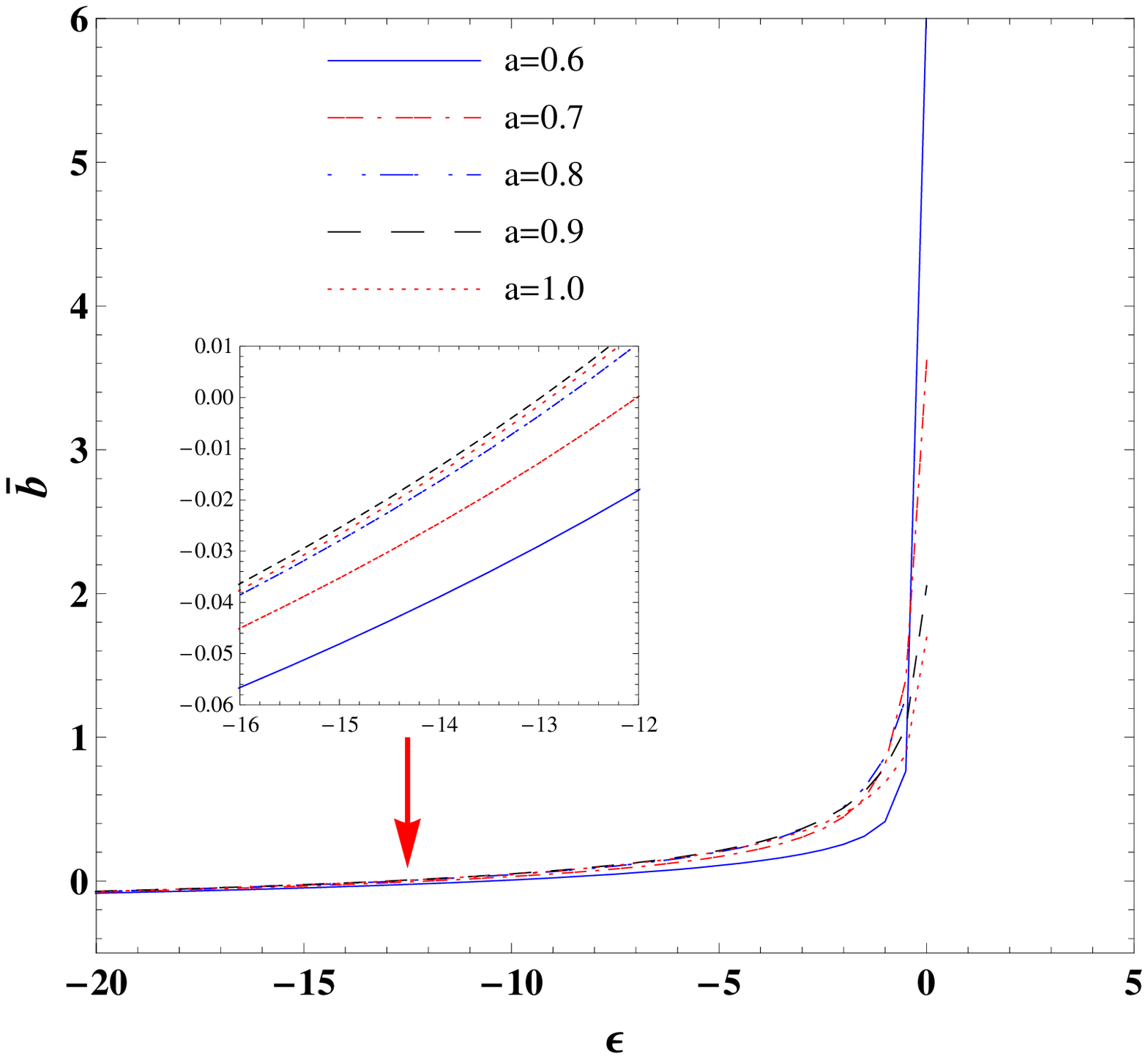} \caption{Change of the
strong deflection limit coefficients with the deformed parameter
$\epsilon$ for different $a$ in the rotating non-Kerr spacetime.
Here, we set $2M=1$.}
\end{center}
\end{figure}
\begin{eqnarray}
z=1-\frac{x_0}{x},
\end{eqnarray}
and rewrite  Eq.(\ref{int1}) as
\begin{eqnarray}
I(x_0)=\int^{1}_{0}R(z,x_0)f(z,x_0)dz,\label{in1}
\end{eqnarray}
with
\begin{eqnarray}
R(z,x_0)&=&\frac{2x^2}{x_0\sqrt{C(z)}}\frac{\sqrt{B(z)|A(x_0)|}[D(z)+JA(z)]}{\sqrt{D^2(z)+A(z)C(z)}},
\end{eqnarray}
\begin{eqnarray}
f(z,x_0)&=&\frac{1}{\sqrt{A(x_0)-A(z)\frac{C(x_0)}{C(z)}+\frac{2J}{C(z)}[A(z)D(x_0)-A(x_0)D(z)]}}.
\end{eqnarray}
\begin{figure}[ht]
\begin{center}
\includegraphics[width=5.0cm]{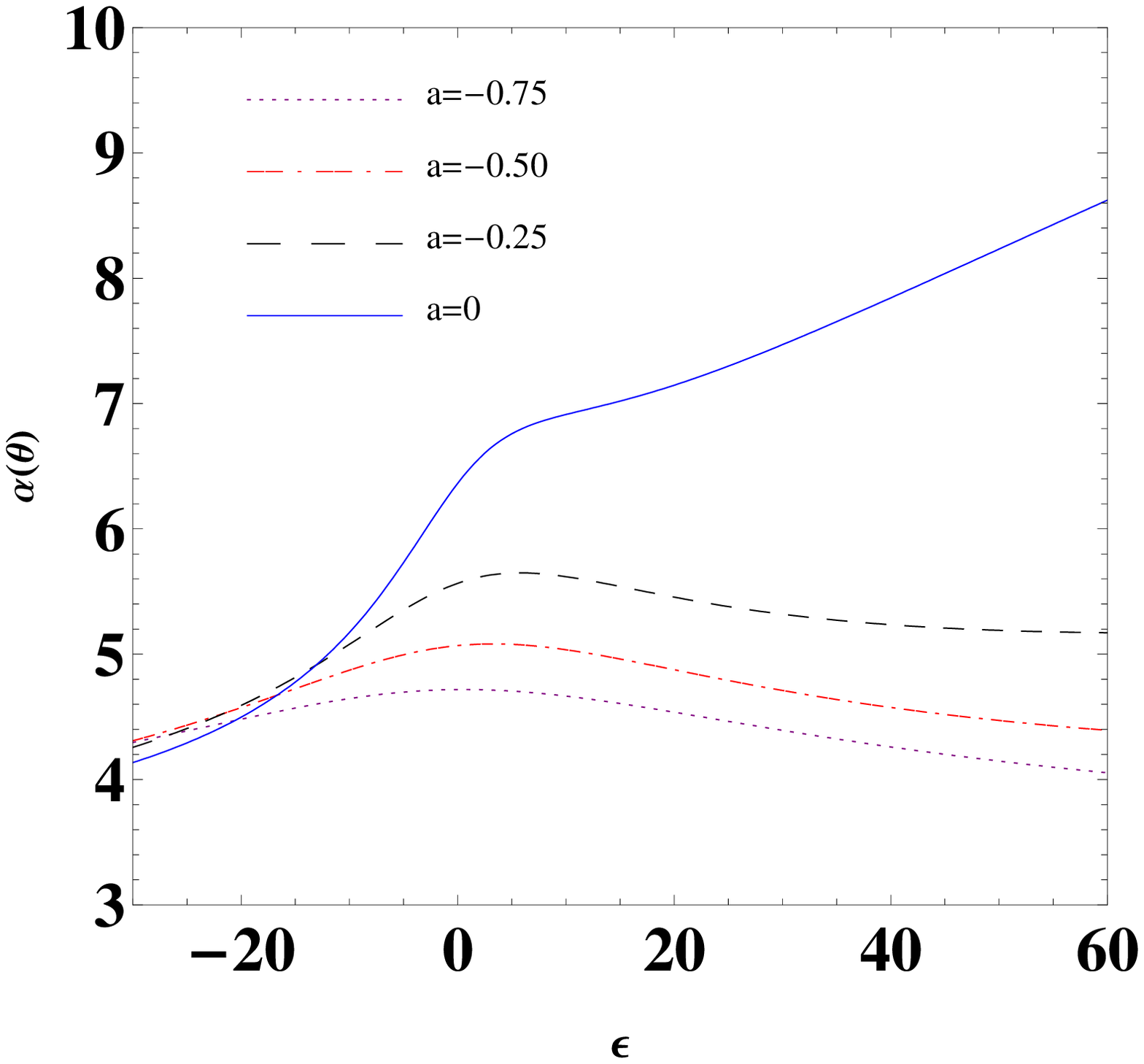}\;\;\includegraphics[width=5.5cm]{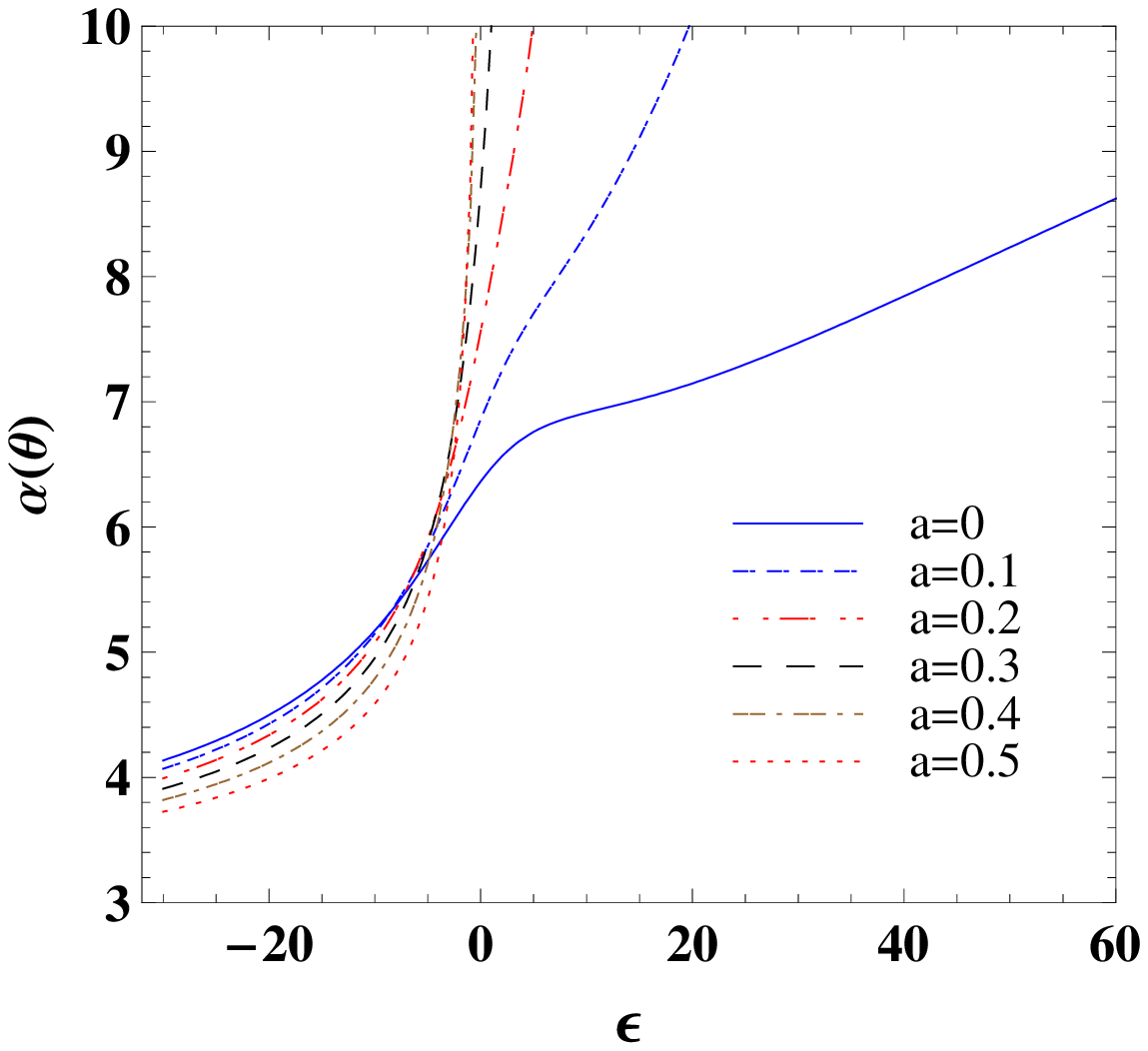}
\;\;\includegraphics[width=5.0cm]{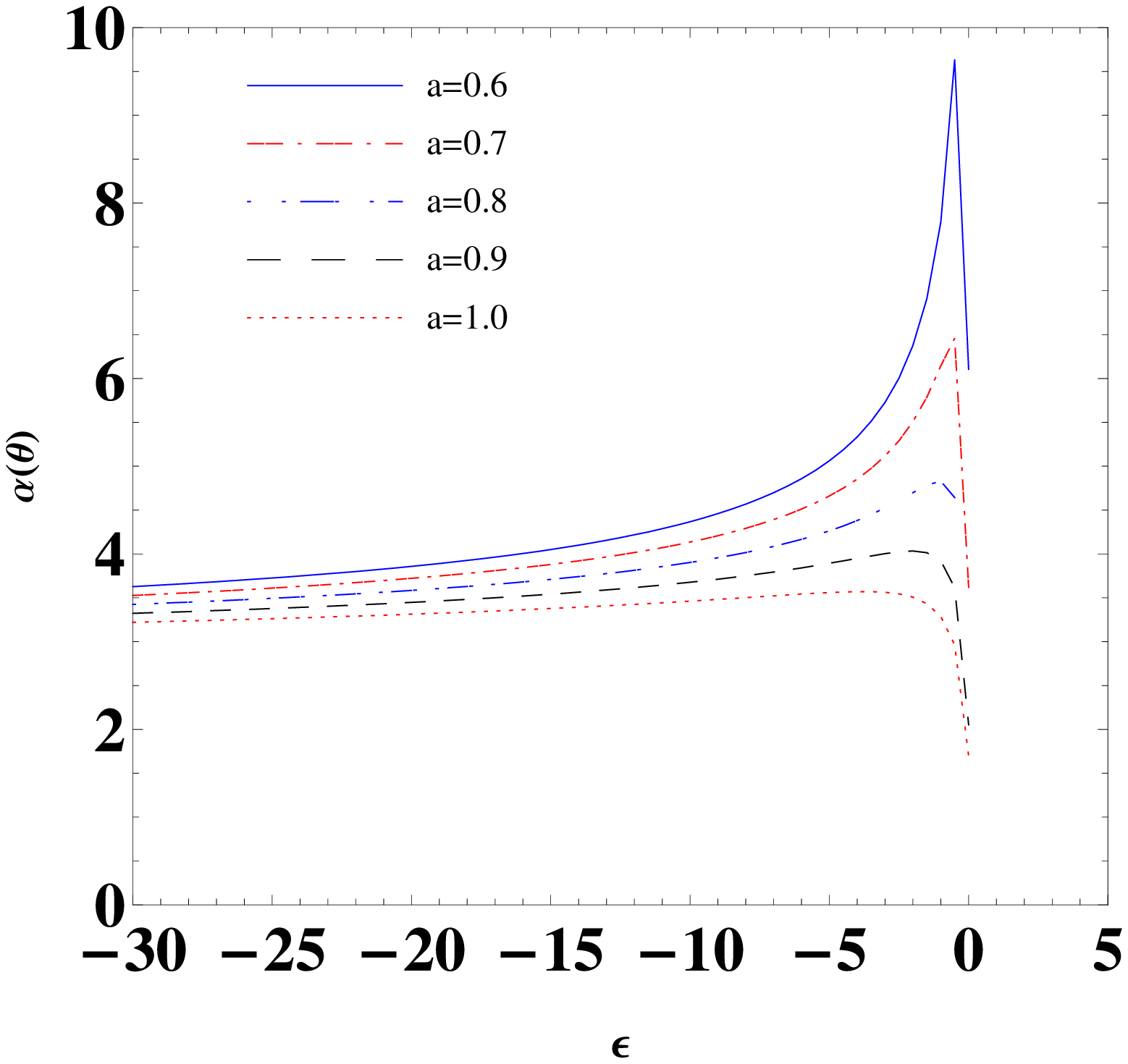}\caption{Deflection
angles evaluated at $u=u_{ps}+0.003$ is a function of the deformed
parameter $\epsilon$ for different $a$. Here, we set $2M=1$.}
\end{center}
\end{figure}
Obviously, the function $R(z, x_0)$ is regular for all values of $z$
and $x_0$. However, the function $f(z, x_0)$ diverges as $z$ tends
to zero, i.e., as the photon approaches the marginally circular
photon orbit. Thus, the integral (\ref{in1}) is separated in two
parts $I_D(x_0)$ and $I_R(x_0)$
\begin{eqnarray}
I_D(x_0)&=&\int^{1}_{0}R(0,x_{ps})f_0(z,x_0)dz, \nonumber\\
I_R(x_0)&=&\int^{1}_{0}[R(z,x_0)f(z,x_0)-R(0,x_{ps})f_0(z,x_0)]dz
\label{intbr}.
\end{eqnarray}
Expanding the argument of the square root in $f(z,x_0)$ to the
second order in $z$, we have
\begin{eqnarray}
f_0(z,x_0)=\frac{1}{\sqrt{p(x_0)z+q(x_0)z^2}},
\end{eqnarray}
where
\begin{eqnarray}
p(x_0)&=&\frac{x_0}{C(x_0)}\bigg\{A(x_0)C'(x_0)-A'(x_0)C(x_0)+2J[A'(x_0)D(x_0)-A(x_0)D'(x_0)]\bigg\},  \nonumber\\
q(x_0)&=&\frac{x_0}{2C^2(x_0)}\bigg\{2\bigg(C(x_0)-x_0C'(x_0)\bigg)\bigg([A(x_0)C'(x_0)-A'(x_0)C(x_0)]
+2J[A'(x_0)D(x_0)-A(x_0)D'(x_0)]\bigg)\nonumber\\&&+x_0C(x_0)
\bigg([A(x_0)C''(x_0)-A''(x_0)C(x_0)]+2J[A''(x_0)D(x_0)-A(x_0)D''(x_0)]\bigg)\bigg\}.\label{al0}
\end{eqnarray}
Comparing Eq.(\ref{root}) with Eq.(\ref{al0}), one can find that if
$x_{0}$ approaches the radius of the marginally circular photon
orbit $x_{ps}$ the coefficient $p(x_{0})$ vanishes and the leading
term of the divergence in $f_0(z,x_{0})$ is $z^{-1}$, which implies
that the integral (\ref{in1}) diverges logarithmically. The
coefficient $q(x_0)$ takes the form
\begin{eqnarray}
q(x_{ps})&=&\frac{x^2_{ps}}{2C(x_{ps})}\bigg\{A(x_{ps})C''(x_{ps})-A''(x_{ps})C(x_{ps})+
2J[A''(x_{ps})D(x_{ps})-A(x_{ps})D''(x_{ps})]\bigg\}.
\end{eqnarray}
Therefore, the deflection angle in the strong field region can be
approximated very well as \cite{Bozza2}
\begin{eqnarray}
\alpha(\theta)=-\bar{a}\log{\bigg(\frac{\theta
D_{OL}}{u_{ps}}-1\bigg)}+\bar{b}+\mathcal{O}(u-u_{ps}), \label{alf1}
\end{eqnarray}
with
\begin{eqnarray}
&\bar{a}&=\frac{R(0,x_{ps})}{\sqrt{q(x_{ps})}}, \nonumber\\
&\bar{b}&= -\pi+b_R+\bar{a}\log{\bigg\{\frac{2q(x_{ps})C(x_{ps})}{u_{ps}A(x_{ps})[D(x_{ps})+JA(x_{ps})]}\bigg\}}, \nonumber\\
&b_R&=I_R(x_{ps}), \nonumber\\
&u_{ps}&=\frac{-D(x_{ps})+\sqrt{A(x_{ps})C(x_{ps})+D^2(x_{ps})}}{A(x_{ps})}.\label{coa1}
\end{eqnarray}
The quantity $D_{OL}$ is the distance between the observer and
the gravitational lens. Making use of Eqs.(\ref{alf1}) and (\ref{coa1}),
we can study the properties of strong gravitational lensing in the
rotating non-Kerr spacetime (\ref{metric0}). In Fig.(6), we plotted
numerically the changes of the coefficients ($\bar{a}$ and $\bar{b}$
) with the deformed parameter $\epsilon$ for different $a$. It is
shown that the coefficients ($\bar{a}$ and $\bar{b}$ ) in the strong
field limit are functions of the rotation parameter $a$ and the
deformed parameter $\epsilon$. For the retrograde photon (i.e.,
$a<0$),  the coefficient $\bar{a}$ first increases up to its maximum
with $\epsilon$, and then decreases down to its minimum with the
further increase of $\epsilon$, after that it increases with
$\epsilon$. The variety of $\bar{b}$ with $\epsilon$ is converse to
the variety of $\bar{a}$ with $\epsilon$ in this case. When $0\leq
a\leq0.5$, $\bar{a}$ increases monotonically with $\epsilon$ and the
behavior of $\bar{b}$ is similar to that in the cases $a<0$. Both
coefficients $\bar{a}$ and $\bar{b}$ diverge as the deformed
parameter $\epsilon$ tends to the upper limit $\epsilon_{max}$ which
still holds up a the marginally circular photon orbit. When $a>0.5$,
coefficients $\bar{a}$ and $\bar{b}$ grow with the increase of
$\epsilon$. As in the vicinity of the upper limit $\epsilon_{max}$
$\bar{a}$ vanishes and $\bar{b}$  tends to a certain positive finite
value, which depends on the parameters $a$ and $\epsilon$. The
divergence of the coefficients of the expansion implies that the
deflection angle in the strong deflection limit (\ref{alf1}) is not
valid in the regime $\epsilon>\epsilon_{max}$. Moreover, we also
note that as the black hole is more oblate, the dependence of the
coefficients $\bar{a}$ and $\bar{b}$ on the rotation parameter $a$
is entirely different from those in the usual Kerr black hole. With
the increase of $a$, the coefficient $\bar{a}$ increases as the
black hole is prolate (i.e., $\epsilon>0$) and decreases  as the
black hole is more oblate. From Fig.(6), one can find that the variety of
$\bar{b}$ with $a$ is converse to the variety of $\bar{a}$ with $a$.
From the above discussion, we know that the change of the
coefficients $\bar{a}$ and $\bar{b}$ become more complicated in the
rotating non-Kerr spacetime (\ref{metric0}). The main reason is that
in this spacetime the presence of the deformation parameter
$\epsilon$ leads to some complicated coupling among the parameters
$a$, $\epsilon$ and the polar coordinate $x$, such as, $a\epsilon$,
$a^2\epsilon$,  $x^{-3}\epsilon$, $x^{2}\epsilon$ and
$a^2x\epsilon$. The effects of these coupling on the coefficients
$\bar{a}$ and $\bar{b}$ are determined  not only by the values of
the parameters $a$ and $\epsilon$, but also by the signs of $a$ and
$\epsilon$. In other words, the dominated coupling terms in the
coefficients $\bar{a}$ and $\bar{b}$ are different at different
regions of values for the parameters $a$ and $\epsilon$, which yields
that $\bar{a}$ and $\bar{b}$ are not the monotonic functions of $a$
and $\epsilon$. The positions of the local minimum and maximum in $\bar{a}$ and $\bar{b}$ are determined by the total effects of these coupling terms and the forms of $\bar{a}$ and $\bar{b}$ themselves.
Furthermore, we plotted in Fig. (7) the change of the deflection
angles evaluated at $u=u_{ps}+0.003$ with $\epsilon$ for different
$a$ as in the regime $\epsilon<\epsilon_{max}$. It is shown that in
the strong field limit the deflection angles have  similar
properties of the coefficient $\bar{a}$. This means that the
deflection angles of the light rays are dominated by the logarithmic
term in this case.

Let us now study the effect of the deformed parameter $\epsilon$
on the observational gravitational lensing parameters in the strong field limit. When the source and observer are far enough from the lens, the lens equation can be approximated well as \cite{Bozza3}
\begin{eqnarray}
\gamma=\frac{D_{OL}+D_{LS}}{D_{LS}}\theta-\alpha(\theta) \; mod
\;2\pi
\end{eqnarray}
where $D_{LS}$ is the lens-source distance and $D_{OL}$ is the
observer-lens distance. $\gamma$ is the angle between the direction
of the source and the optical axis. $\theta=u/D_{OL}$ is the angular
separation between the lens and the image. Following
Ref.\cite{Bozza3},  we here
consider only the case in which the source, lens and observer are
highly aligned. In this limit, one can find that the angular separation between the lens and the n-th relativistic image is
\begin{eqnarray}
\theta_n\simeq\theta^0_n\bigg(1-\frac{u_{ps}e_n(D_{OL}+D_{LS})}{\bar{a}D_{OL}D_{LS}}\bigg),
\end{eqnarray}
with
\begin{eqnarray}
\theta^0_n=\frac{u_{ps}}{D_{OL}}(1+e_n),\;\;\;\;\;\;e_{n}=e^{\frac{\bar{b}+|\gamma|-2\pi
n}{\bar{a}}}.
\end{eqnarray}
\begin{figure}[ht]
\begin{center}
\includegraphics[width=5.2cm]{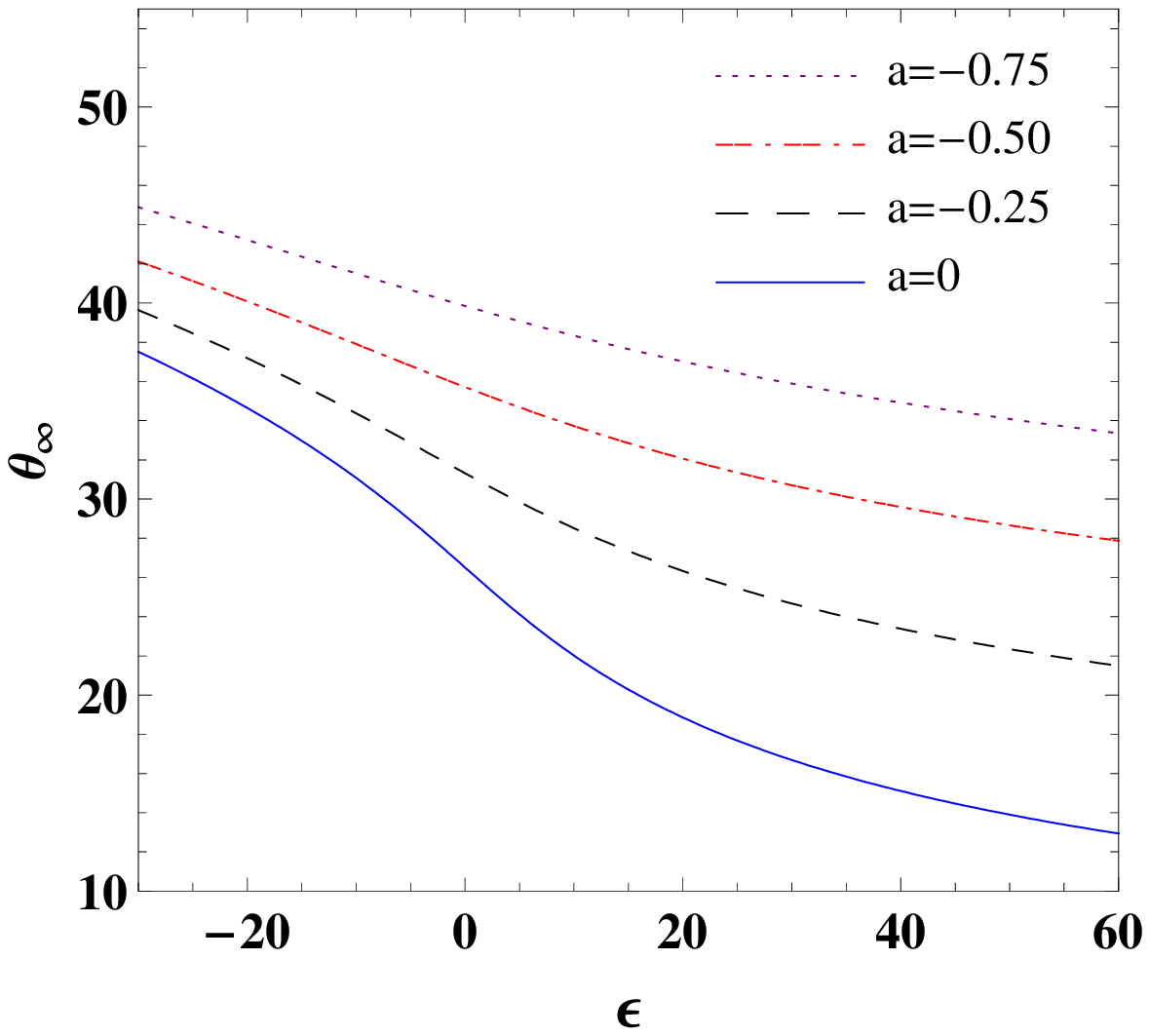}\;\;\includegraphics[width=5.5cm]{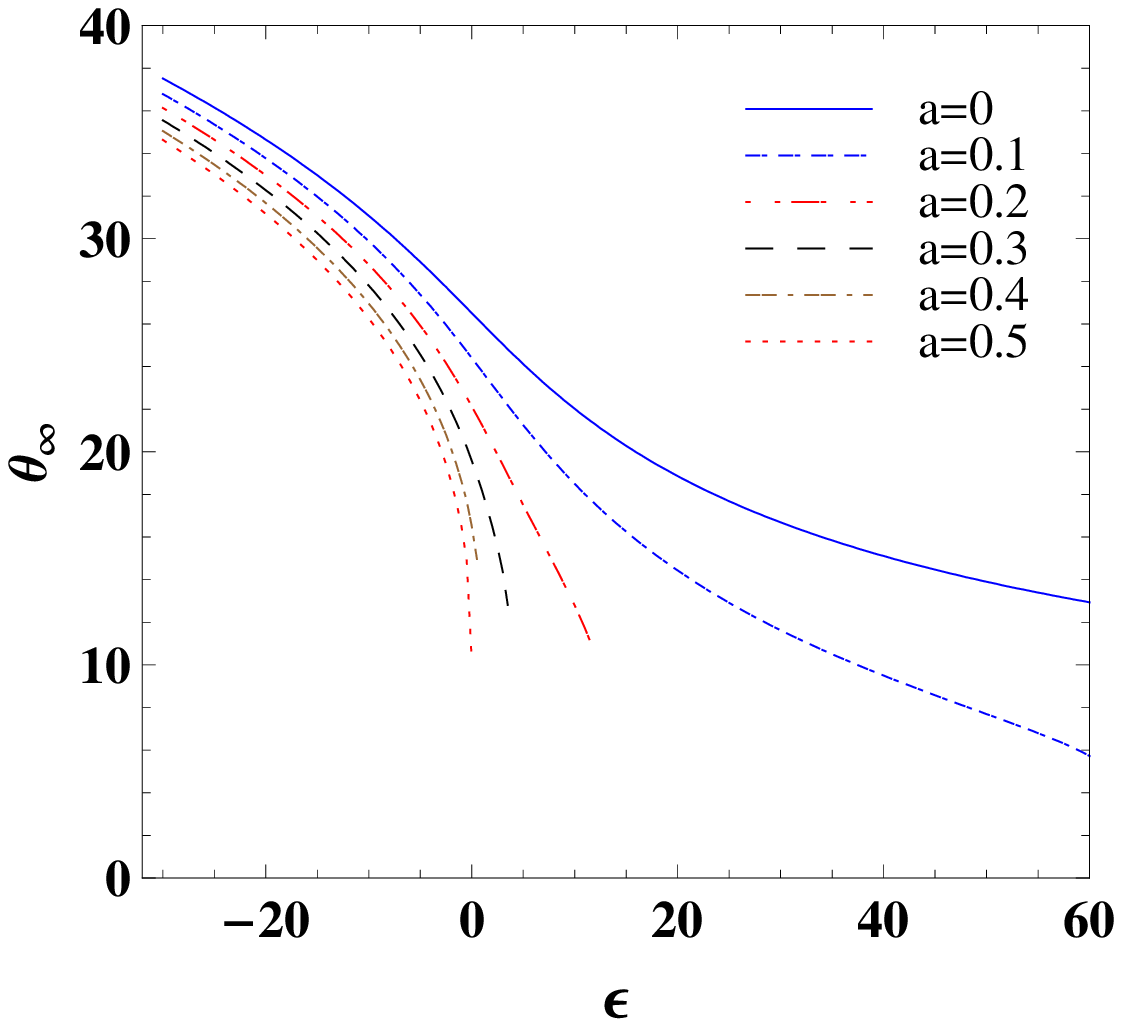}
\;\;\includegraphics[width=5.5cm]{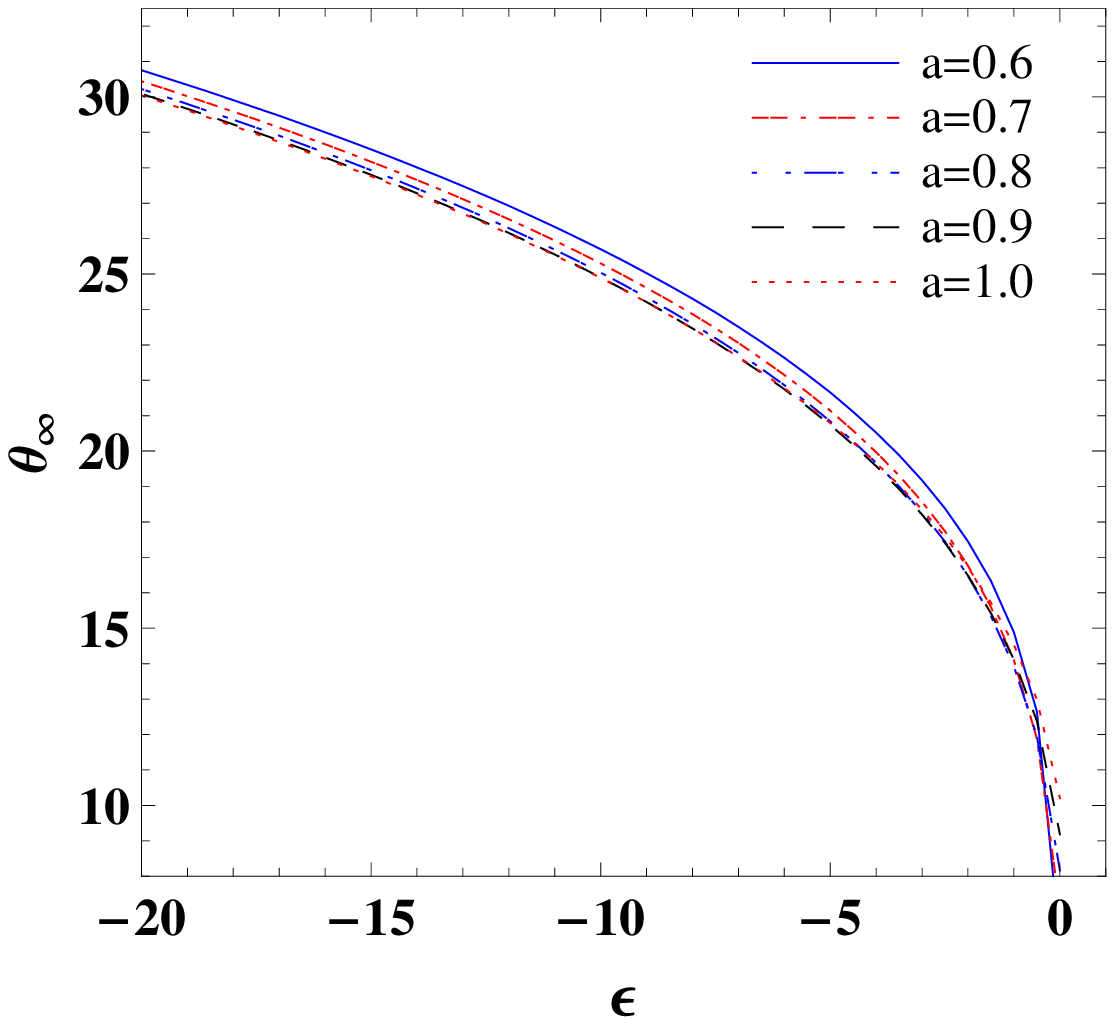}\caption{Variety of the
innermost relativistic image $\theta_{\infty}$ with the deformed
parameter $\epsilon$ for different $a$. Here, we set $2M=1$.}
\end{center}
\end{figure}
\begin{figure}[ht]
\begin{center}
\includegraphics[width=5.2cm]{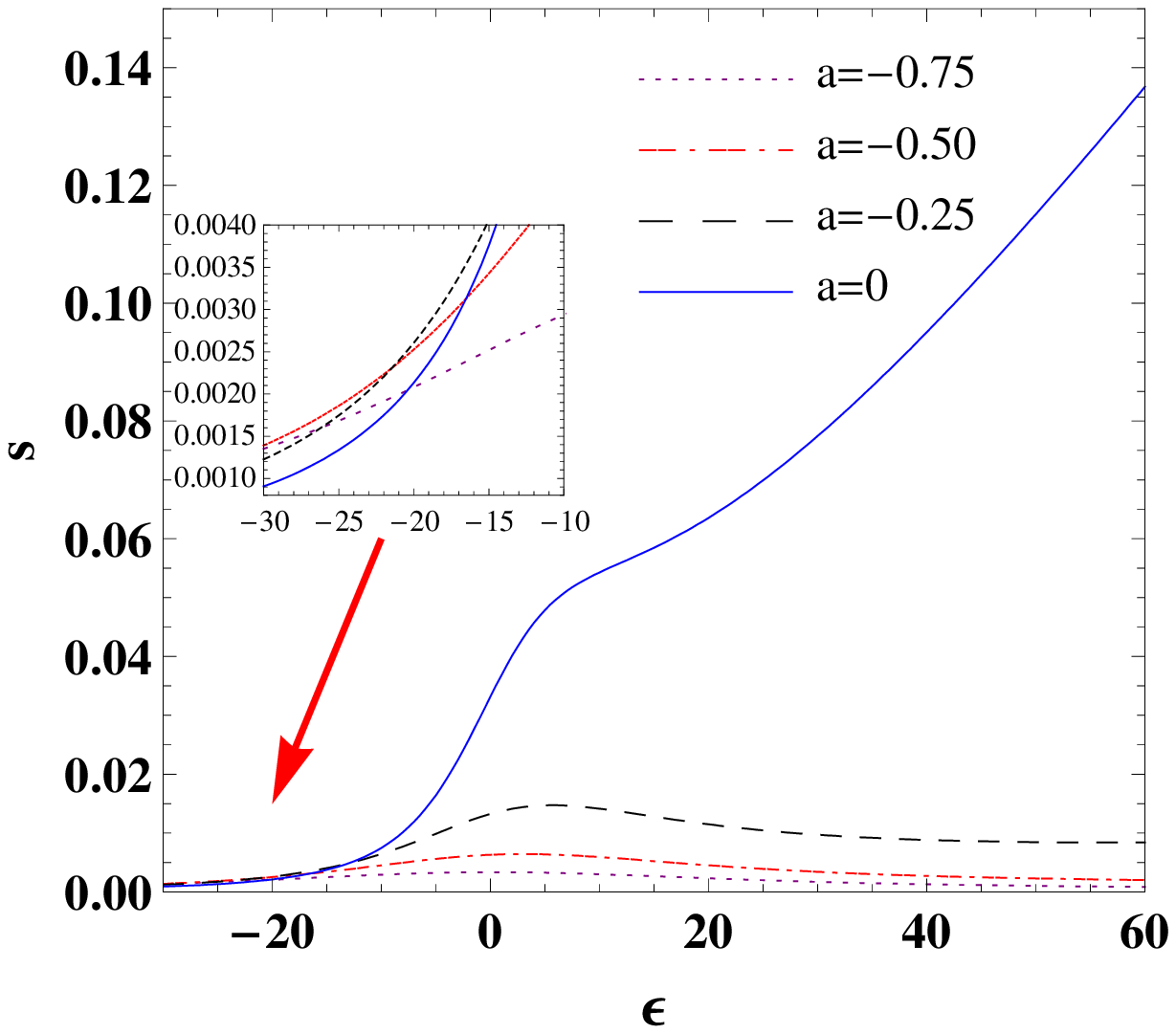}\;\;\includegraphics[width=5.2cm]{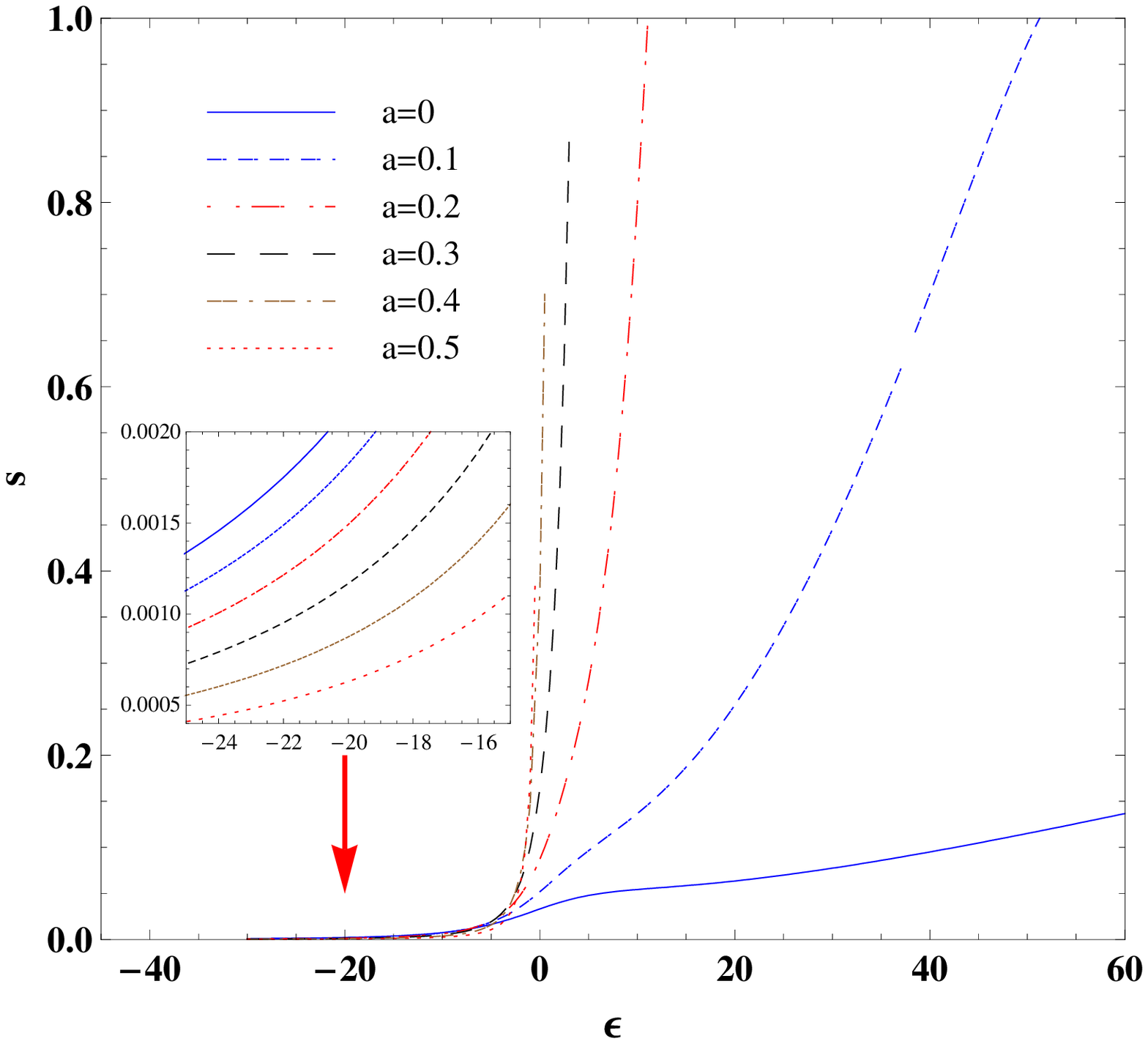}
\;\;\includegraphics[width=5.2cm]{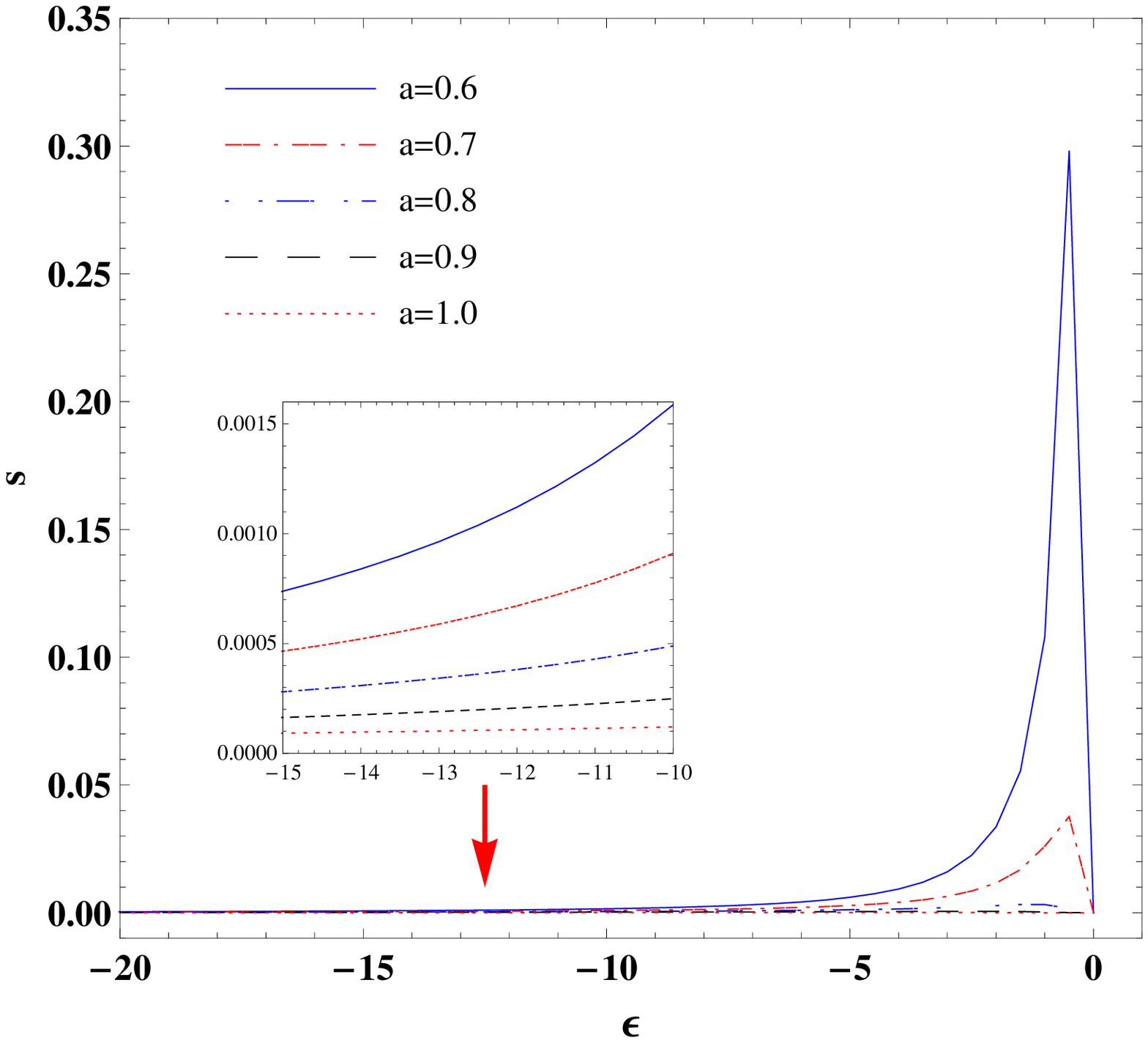}\caption{Variety of the
angular separation $s$ with the deformed parameter $\epsilon$ for
different $a$. Here, we set $2M=1$.}
\end{center}
\end{figure}
\begin{figure}[ht]
\begin{center}
\includegraphics[width=5.8cm]{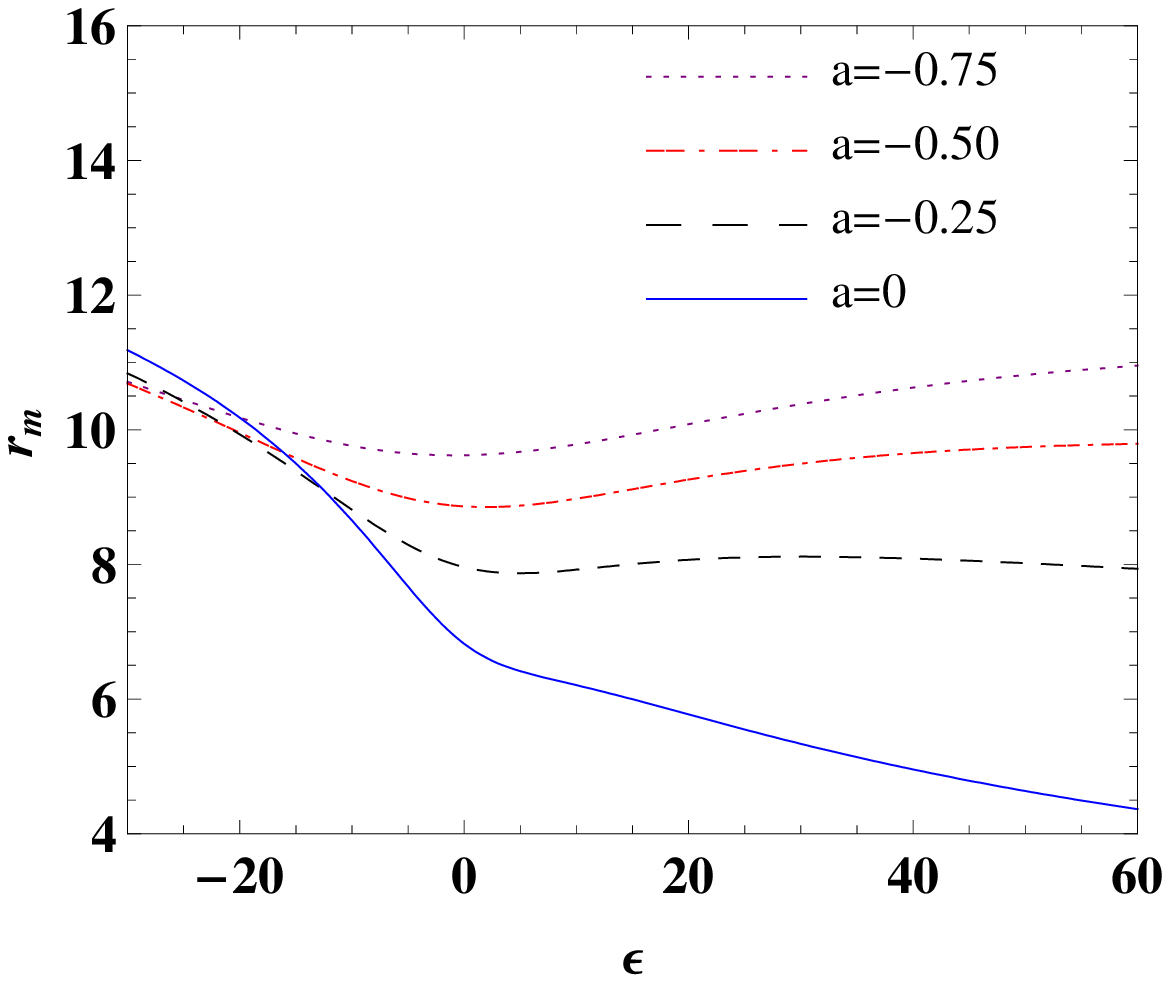}\;\;\includegraphics[width=5.2cm]{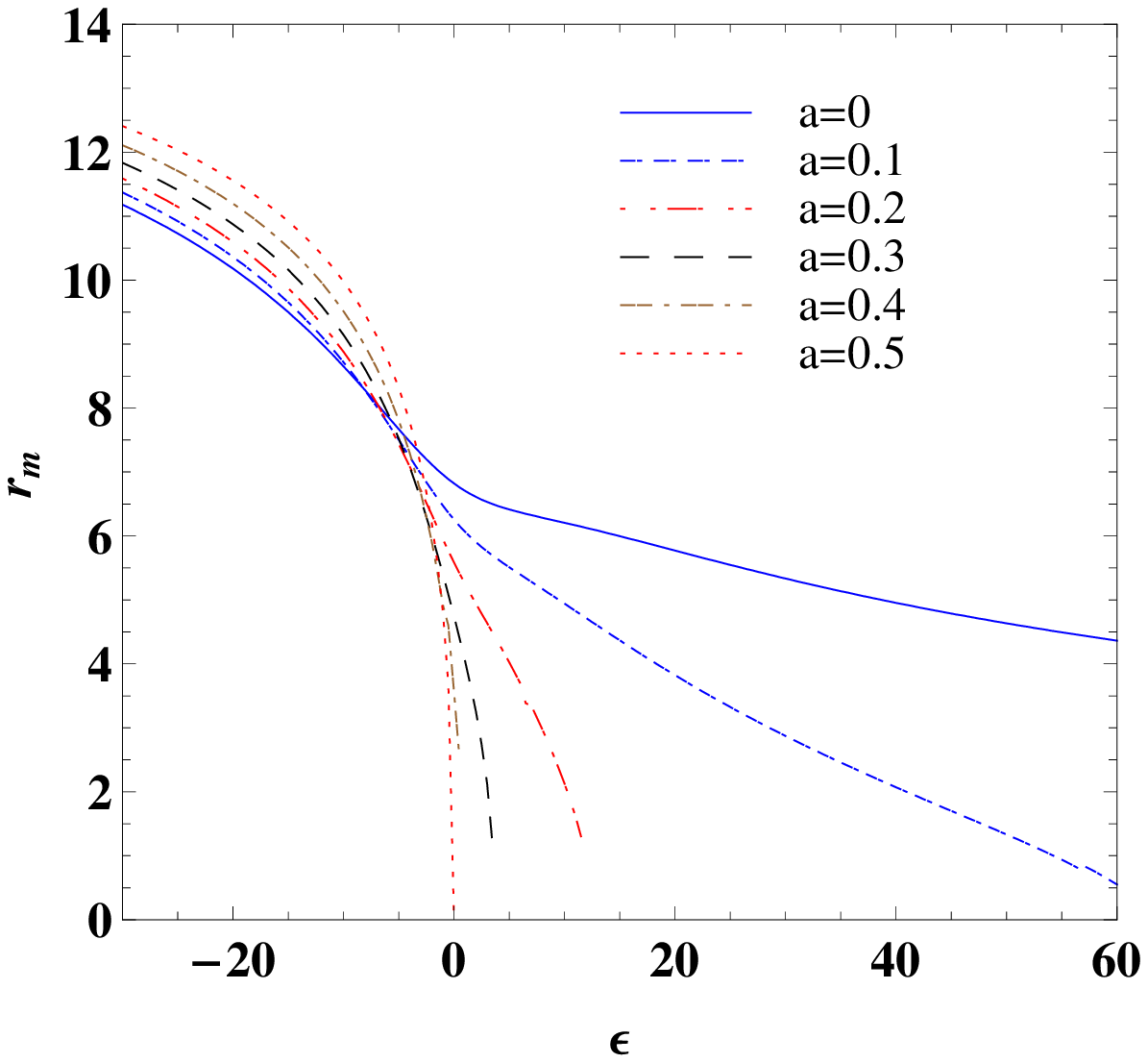}
\;\;\includegraphics[width=5.2cm]{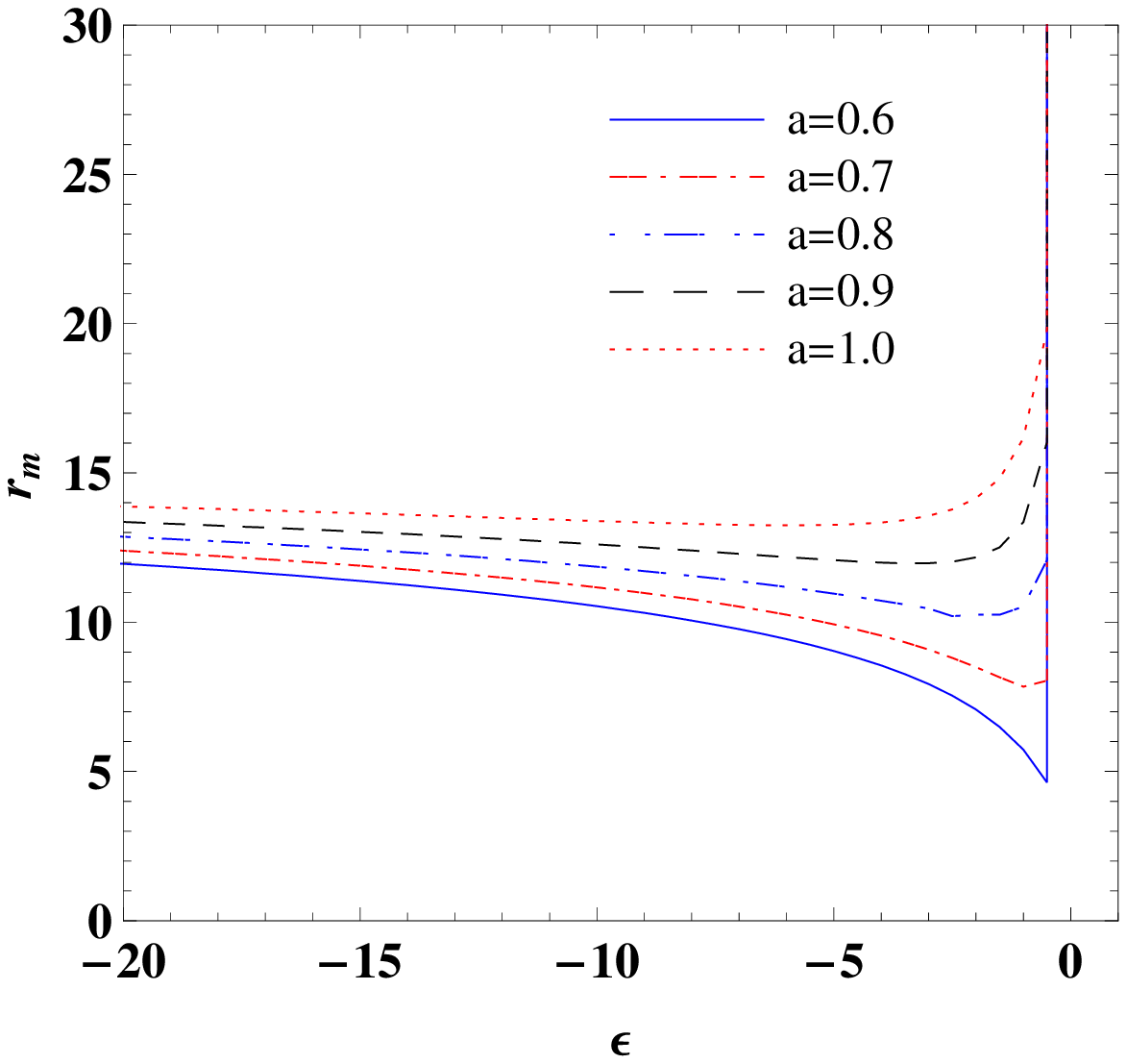}\caption{Variety of the
relative magnitudes $r_m$ with the deformed parameter $\epsilon$ for
different $a$. Here, we set $2M=1$.}
\end{center}
\end{figure}
The quantity $\theta^0_n$ is the image position corresponding to
$\alpha=2n\pi$, and $n$ is an integer. According to the past
oriented light ray which starts from the observer and finishes at
the source the resulting images stand on the eastern side of the
black hole for direct photons ($a>0$) and are described by positive
$\gamma$. Retrograde photons ($a<0$) have images on the western side
of the black hole and are described by negative values of $\gamma$.
In the limit $n\rightarrow \infty$, one can find that
$e_n\rightarrow 0$, which means that the relation between the
minimum impact parameter $u_{ps}$ and the asymptotic position of a
set of images $\theta_{\infty}$ can be simplified further as
\begin{eqnarray}
u_{ps}=D_{OL}\theta_{\infty}.\label{uhs1}
\end{eqnarray}
In order to obtain the coefficients $\bar{a}$ and $\bar{b}$, one
needs to separate at least the outermost image from all the others.
As in Refs.\cite{Bozza2,Bozza3},  we consider here the simplest case
in which only the outermost image $\theta_1$ is resolved as a single
image and all the remaining ones are packed together at
$\theta_{\infty}$. Thus the angular separation $s$ between the first
image and the other ones and the ratio of the flux from the first image
and those from all other images $\mathcal{R}$ can be expressed
as \cite{Bozza2,Bozza3,Gyulchev1}
\begin{eqnarray}
s&=&\theta_1-\theta_{\infty}=\theta_{\infty}e^{\frac{\bar{b}-2\pi}{\bar{a}}},\nonumber\\
\mathcal{R}&=&\frac{\mu_1}{\sum^{\infty}_{n=2}\mu_n}=e^{\frac{2\pi}{\bar{a}}}.\label{ss1}
\end{eqnarray}
Through measuring $s$, $\theta_{\infty}$ and $\mathcal{R}$, we can
obtain the strong deflection limit coefficients $\bar{a}$ and $\bar{b}$
and the minimum impact parameter $u_{ps}$. Comparing their values
with those predicted by theoretical models, we can obtain
information about the parameters of the lens object stored in them.

The mass of the central object of our Galaxy was estimated recently
to be $4.4\times 10^6M_{\odot}$ \cite{Genzel1} and its distance is
around $8.5kpc$, so that the ratio of the mass to the distance
$M/D_{OL} \approx2.4734\times10^{-11}$.  Making use of Eqs.
(\ref{coa1}), (\ref{uhs1}) and  (\ref{ss1})  we can estimate the
values of the coefficients and observables for gravitational lensing
in the strong field limit. The numerical value for the angular
position of the relativistic images $\theta_{\infty}$, the angular
separation $s$ and the relative magnitudes $r_m$ (which is related
to $\mathcal{R}$ by $r_m=2.5\log{\mathcal{R}}$) are plotted in
Figs.(8), (9) and (10). In generally, the angular position of the
relativistic images $\theta_{\infty}$ decreases with the parameters
$\epsilon$ and $a$. While in the vicinity $\epsilon=0$, it increases
with the rotation parameter $a$ as $a>0.5$. For the retrograde
photon (i.e., $a<0$), the angular separation $s$ between
$\theta_{1}$ and $\theta_{\infty}$ first increases up to its maximum
with $\epsilon$, and then decreases down to its minimum with the
further increase of $\epsilon$, after that it increases with
$\epsilon$. For the prograde photon (i.e., $a>0$), $s$ increases
with $\epsilon$, but near the region $\epsilon\sim0$ it decreases
with $\epsilon$ and finally tends to zero as $a>0.5$. The change of
$r_m$ with the parameters $\epsilon$ and $a$ is converse to that of
the coefficient $\bar{a}$. Comparing with those in the Kerr
background, one find that the behavior of these observable become
more complicated since these observable are the functions of the
coefficients $\bar{a}$ and $\bar{b}$, which are related to the
parameters $a$ and $\epsilon$ with some complex forms. The intrinsic
reason is that the presence of $\epsilon$ changes the structure of
spacetime and makes the motion of the photon more complicated.

\section{summary}

In this paper we have investigated the features of gravitational
lensing in a four-dimensional rotating non-Kerr spacetime proposed
recently by Johannsen and Psaltis \cite{TJo} to test the no-hair
theorem. Our results show that the deformed parameter $\epsilon$ and
the rotation parameter $a$ imprint in the marginally circular photon
orbit, the deflection angle, the coefficients in strong field
lensing and the observational gravitational lensing variables. We
find that the marginally circular photon orbit radius $x_{ps}$
always exists for the cases $\epsilon<0$ or  $a<0$. But for the case
$\epsilon>0$ and $a>0$, it exists only in the regime
$\epsilon\leq\epsilon_{max}$ for fixed $a$. The upper limit
$\epsilon_{max}$ is defined by the critical condition that the
marginally circular photon orbit is overlapped with the event
horizon. As $\epsilon>\epsilon_{max}$, both the marginally circular
photon orbit and the event horizon vanish and then the singularity
is naked. The deflection angle of the light ray very close to the
naked singularity is a positive finite value, which is different
from those in the usual Kerr black hole spacetime and in the
rotating naked singularity described by the Janis-Newman-Winicour
metric. For all values of $a$, the relativistic images are closer to
the optical axis for a larger deformed parameter $\epsilon$. When
$a<0$, the separability $s$ first increases up to its maximum with
$\epsilon$, and then decreases down to its minimum with the further
increase of $\epsilon$, after that it increases with $\epsilon$.
When $a>0$, $s$ increases with $\epsilon$, but near the region
$\epsilon\sim0$ it decreases with $\epsilon$ and finally tends to
zero as $a>0.5$. The change of $r_m$ with the parameters $\epsilon$
and $a$ is converse to that of the coefficient $\bar{a}$. Moreover,
we also note that as the black hole is more oblate, the dependence
of the coefficients $\bar{a}$, $\bar{b}$ and the separability $s$ on
the rotation parameter $a$ is entirely different from those in the
usual Kerr black hole. These significant features, at least in
principle, may provide a possibility to test the no-hair theorem in
future astronomical observations.

\section{\bf Acknowledgments}

This work was  partially supported by the NCET under Grant
No.10-0165, the PCSIRT under Grant No. IRT0964 and the construct
program of key disciplines in Hunan Province. J. Jing's work was
partially supported by the National Natural Science Foundation of
China under Grant Nos. 11175065, 10935013; 973 Program Grant No.
2010CB833004.


\vspace*{0.2cm}
\begin{thebibliography}{99}

\baselineskip=0.6 cm \baselineskip=0.6 cm

\bibitem{noh} W. Israel, Phys. Rev. {\bf164} 1776 (1967) ;
W. Israel, Commun. Math. Phys. {\bf8} 245 (1968); B. Carter, Phys.
Rev. Lett. {\bf26} (1971) 331; S. W. Hawking, Commun. Math. Phys.
{\bf25} (1972) 152; D. C. Robinson, Phys. Rev. Lett. {\bf34} 905
(1975).

\bibitem{gwave} F. D. Ryan, Phys. Rev. D {\bf52}  5707 (1995); L. Barack
and C. Cutler, Phys. Rev. D {\bf69} 082005 (2004); J. Brink, Phys.
Rev. D {\bf78} 102001 (2008); C. Li and G. Lovelace, Phys. Rev. D
{bf77} 064022 (2008); T. A. Apostolatos, G. Lukes-Gerakopoulos, and
G. Contopoulos, Phys. Rev. Lett. {\bf103}  111101 (2009).

\bibitem{ga1} N. A. Collins and S. A. Hughes,
Phys. Rev. D {\bf69} 124022 (2004).

\bibitem{ga2} S. J. Vigeland and S. A. Hughes, Phys. Rev. D {\bf81}
024030 (2010).

\bibitem{JGa}  J. Gair, C. Li and I. Mandel, Phys. Rev. D {\bf77}
024035 (2008).

\bibitem{ag1} T. Johannsen and D. Psaltis, Adv. Space Res. {\bf47}
528 (2011).


\bibitem{CBa3} C. Bambi and E. Barausse, Astrophys. J. {\bf731}
121 (2011).

\bibitem{ag3} V. S. Manko and I. D. Novikov, Class. Quantum Grav. {\bf9} 2477 (1992).

\bibitem{ag4} S. J. Vigeland, N. Yunes, and L. C. Stein, Phys. Rev. D {\bf} 83
104027 (2011).




\bibitem{TJo} T. Johannsen, D. Psaltis, Phys. Rev. D {\bf83} 124015 (2011).

\bibitem{JNT} E. T. Newman and A. I. Janis, J. Math. Phys. {\bf6} 915 (1965);
S. P. Drake and P. Szekeres, Gen. Rel. Grav. {\bf32} 445 (2000).

\bibitem{CBa} C. Bambi, L. Modesto, Phys. Lett. B {\bf706} 13 (2011).

\bibitem{CBa1} C. Bambi,  Phys. Lett. B {\bf705}  5 (2011);
C. Bambi, L. Modesto, Phys. Lett. B {\bf711}, 10 (2012).

\bibitem{FCa}F. Caravelli, L. Modesto, Class. Quant. Grav. {\bf27}
245022 (2010).

\bibitem{VCa1} P. Pani, C. F. B. Macedo, L. C. B. Crispino, V.
Cardoso,  Phys. Rev. D {\bf84} 087501  (2011).

\bibitem{TJo1} T. Johannsen, Adv. Astron.{\bf2012}, 486750 (2012), arXiv:1105.5645 [astro-ph.HE].

\bibitem{sc} S. Chen, J. Jing, Phys. Lett. B {\bf711} 81 (2012), arXiv:1110.3462.


\bibitem{Darwin} C. Darwin, Proc. of the Royal Soc. of London {\bf 249}
180 (1959).

\bibitem{Vir} K. S. Virbhadra, D. Narasimha and S. M. Chitre,
Astron. Astrophys. {\bf 337} 1 (1998).

\bibitem{Vir1} K. S. Virbhadra, G. F. R. Ellis,  Phys. Rev. D {\bf 62}
084003 (2000).

\bibitem{Vir2} C. M. Claudel, K. S. Virbhadra, G. F. R. Ellis,
J. Math. Phys. {\bf 42} 818 (2001).

\bibitem{Vir3} K. S. Virbhadra, G. F. R. Ellis, Phys. Rev.D {\bf 65}
103004 (2002).


\bibitem{Bozza2} V. Bozza, Phys. Rev. D {\bf 66} 103001 (2002).

\bibitem{Bozza3} V. Bozza, Phys. Rev. D {\bf 67} 103006 (2003);

V. Bozza, F. De Luca, G. Scarpetta, M. Sereno,  Phys. Rev. D {\bf
72} 083003 (2005);

V. Bozza, F. De Luca, G. Scarpetta,  Phys. Rev. D {\bf 74} 063001
(2006).





\bibitem{Gyulchev} G. N. Gyulchev and S. S. Yazadjiev, Phys. Rev. D {\bf75}
023006 (2007).


\bibitem{Gyulchev1} G. N. Gyulchev and S. S. Yazadjiev, Phys. Rev. D {\bf78}
083004 (2008).


\bibitem{Fritt} S. Frittelly, T. P. Kling, E. T. Newman, Phys. Rev. D {\bf 61}
064021 (2000).

\bibitem{Bozza1}V. Bozza, S. Capozziello, G. lovane, G.
Scarpetta, Gen. Rel. and Grav. {\bf 33} 1535 (2001).

\bibitem{Eirc1}E. F. Eiroa,  G. E. Romero, D. F. Torres, Phys. Rev. D {\bf 66}
024010 (2002).

E. F. Eiroa,  Phys. Rev. D {\bf 71} 083010 (2005);

E. F. Eiroa,  Phys. Rev. D {\bf 73} 043002 (2006).

\bibitem{whisk} R. Whisker, Phys. Rev. D {\bf71} 064004 (2005).

\bibitem{Bhad1} A. Bhadra, Phys. Rev. D {\bf 67} 103009 (2003).

\bibitem{Song1} S. Chen and J. Jing,  Phys. Rev. D {\bf 80}
024036 (2009).


\bibitem{Song2} Y. Liu,  S. Chen and J. Jing, Phys. Rev. D {\bf81}
124017 (2010);  S. Chen, Y. Liu  and J. Jing, Phys. Rev. D {\bf83}
124019 (2011).

\bibitem{TSa1} T. Ghosh, S. Sengupta, Phys. Rev. D {\bf81} 044013
(2010).

\bibitem{AnAv} A. N. Aliev, P. Talazan,  Phys. Rev. D {\bf80} 044023
(2009).

\bibitem{Ls1} S. Wei, Y. Liu, Phys. Rev. D {\bf85}, 064044 (2012).



\bibitem{Kraniotis} G. V. Kraniotis, Class. Quant. Grav. {\bf28}, 085021
(2011).


\bibitem{Ein1} A. Einstein, Science {\bf84} 506 (1936).

\bibitem{Genzel1} R. Genzel, F. Eisenhauer, S. Gillessen,
Rev. Mod. Phys. {\bf82}, 3121 (2010), arXiv:1006.0064

\end{thebibliography}
\end{document}